\documentclass[journal]{IEEEtran}
\usepackage{xcolor,soul,framed} %,caption
\usepackage{cite}
\usepackage{amsmath,amssymb,amsfonts}
\usepackage{graphicx}
\usepackage{textcomp}
\usepackage{dcolumn}% Align table columns on decimal point
\usepackage{bm}% bold math
\usepackage{subfigure} %Para poder hacer subplots
\usepackage{pict2e}
\usepackage{epstopdf}
\usepackage{arydshln}
\usepackage{gensymb}
\usepackage{enumitem}
\usepackage[mathscr]{euscript}
\usepackage{stfloats}
\usepackage{diagbox}

\definecolor{bluerevision}{RGB}{0,112,192}
\newcommand{\blue}[1]{\textcolor{black}{#1}}

\begin{document}

\thispagestyle{empty}
\onecolumn

%\noindent\textbf{Copyright}: Copyright: ©2022 IEEE. Personal use of this material is permitted. Permission from IEEE must be obtained for all other uses, in any current or future media, including reprinting/republishing this material for advertising or promotional purposes, creating new collective works, for resale or redistribution to servers or lists, or reuse of any copyrighted component of this work in other works.\\

\noindent\textbf{Disclaimer}: This work has been published in \textit{IEEE Open Journal of the Communications Society}.\\

\noindent Citation information: DOI 10.1109/OJCOMS.2022.3156473

\twocolumn
\setcounter{page}{1}
\newpage

\title{Artificial Intelligence and Dimensionality Reduction: Tools for approaching future communications}

\author{Alejandro Ramírez-Arroyo, Luz García, Antonio Alex-Amor, and Juan F. Valenzuela-Valdés % <-this % stops a space
\thanks{This work was supported in part by the Spanish Program of Research, Development and Innovation under Project RTI2018-102002-A-I00 and, in part by “Junta de Andalucía” under Project B-TIC-402-UGR18 and Project P18.RT.4830 and in part by the predoctoral grant FPU19/01251. (\textit{Corresponding author: Alejandro Ramírez-Arroyo}).}
\thanks{A. Ramírez-Arroyo, L. García and J. F. Valenzuela-Valdés are with the Department of Signal Theory, Telematics and Communications, Universidad de Granada (UGR), 18071 Granada, Spain (e-mail: alera@ugr.es, luzgm@ugr.es, juanvalenzuela@ugr.es).}
\thanks{A. Alex-Amor is with the Information Technologies Department, Universidad CEU San Pablo (USP CEU), 28668 Boadilla del Monte, Madrid, Spain (e-mail: antonio.alexamor@ceu.es).}
}

\markboth{Ramírez-Arroyo \MakeLowercase{\textit{et al.}}: Artificial Intelligence and Dimensionality Reduction: Tools for approaching future communications}%
{Ramírez-Arroyo \MakeLowercase{\textit{et al.}}: Artificial Intelligence and Dimensionality Reduction: Tools for approaching future communications}

\IEEEpeerreviewmaketitle

\maketitle

% As a general rule, do not put math, special symbols or citations in the abstract or keywords.

\begin{abstract}
This article presents a novel application of the t-distributed Stochastic Neighbor Embedding (t-SNE) clustering algorithm to the telecommunication field. t-SNE is a dimensionality reduction algorithm that allows the visualization of large dataset into a 2D plot. We present the applicability of this algorithm in a communication channel dataset formed by several scenarios (anechoic, reverberation, indoor and outdoor), and by using six channel features. Applying this artificial intelligence (AI) technique, we are able to separate different environments into several clusters allowing a clear visualization of the scenarios. Throughout the article, it is proved that t-SNE has the ability to cluster into several subclasses, obtaining internal classifications within the scenarios themselves. t-SNE comparison with different dimensionality reduction techniques (PCA, Isomap) is also provided throughout the paper. Furthermore, post-processing techniques are used to modify communication scenarios, recreating a real communication scenario from measurements acquired in an anechoic chamber. The dimensionality reduction and classification by using t-SNE and Variational AutoEncoders show good performance distinguishing between the recreation and the real communication scenario. The combination of these two techniques opens up the possibility for new scenario recreations for future mobile communications. This work shows the potential of AI as a powerful tool for clustering, classification and generation of new 5G propagation scenarios.
\end{abstract}


\newcommand*{\bigs}[1]{\vcenter{\hbox{\scalebox{2}[8.2]{\ensuremath#1}}}}

\newcommand*{\bigstwo}[1]{\vcenter{\hbox{ \scalebox{1}[4.4]{\ensuremath#1}}}}

%%%%%%%%%%%%%%%%%%%%%%%%%%%%%%%%%%%%%%%%%%%%%%%%%%%%%%%%
\section{Introduction}
%%%%%%%%%%%%%%%%%%%%%%%%%%%%%%%%%%%%%%%%%%%%%%%%%%%%%%%%

\IEEEPARstart{T}{he} growth in wireless communication networks in recent years has been exponential. An increasingly interconnected world, together with technological advances, makes this field one of the main topics in the research community \cite{survey_5G}-\hspace{1sp}\cite{ericsson_report}. 5G emergence promises multiple improvements at the user level, including improved transmission rates, reduced end-to-end delay, reduced power consumption, improved energy efficiency and ultra-densified networks \cite{URLLC}-\hspace{1sp}\cite{network_densification}. In order to provide all these benefits, a deep analysis must be performed over the communication channels considered for these wireless communications. \blue{As user demands are increasing, the diversity of communication scenarios do it as well}. New environments are emerging, as for example, Vehicle-to-Vehicle (V2V) \cite{V2E_survey}, UAV-to-UAV \cite{UAV2UAV}, Ship-to-Ship (S2S) \cite{IoS_survey}, High Speed Train-to-High Speed Train \cite{HST2HST}, or any combination of the above \cite{Air2Ground}. The characterization of these environments will be fundamental to determine the communication channels key performance indicators (KPIs): data rate, reliability, latency or transmit power.

Since the number of these new scenarios is increasing, the complexity of their analysis is also escalating \cite{Multilayer}. In order to solve this problem, Artificial Intelligence (AI) appears as a tool that can be applied in the telecommunication field. Several examples show the feasibility of its use in this field, such as deep learning for microwave imaging \cite{microwave_imaging} and inverse scattering \cite{inverse_scattering}, support vector regression for antenna design \cite{SVR_Arrebola} or deep neural networks for estimation of the Direction-of-Arrival (DoA) \cite{DoA_estimation}. The AI field includes Machine Learning (ML), where classification, clustering and dimensionality reduction (DR) algorithms are found. Multiple classification algorithms have been developed over the last decade, like Deep Convolutional Neural Networks (DCNN) \cite{ImageNet} or Variational AutoEncoders (VAEs) \cite{VAEs}. In the same way, several clustering methods have been developed, for instance, k-means \cite{kmeans} \blue{or} Density-Based Spatial Clustering of Applications with Noise (DBSCAN) \cite{DBSCAN}. Finally, some examples of dimensionality reduction techniques include Principal Component Analysis (PCA) \cite{PCA}, Isometric feature mapping (Isomap) \cite{Isomap}, t-distributed Stochastic Neighbor Embedding (t-SNE) \cite{tSNE} or \blue{Uniform Manifold Approximation and Projection (UMAP) \cite{UMAP}}.

Among the works that combine communication scenarios and AI techniques, the following are worth mentioning:

\begin{itemize}
  \item Zhou et al. \cite{HST_analysis} propose a deep neural network (DNN) and a score fusion scheme for classification purposes. Four scenarios related to high-speed railway channels (Rural, Station, Suburban and Multi-link) are classified  by using four channel features (K Factor, RMS delay spread, RMS Doppler power spectrum and RMS angular spread).
  
  \item Zhang et al. \cite{scenario_identification}  apply several classifiers and clustering techniques [k-nearest neighbor (k-NN), support vector machine (SVM), k-means, and Gaussian mixture model (GMM)] to identify  four simulated scenarios (Urban Macrocell (UMa) and Rural Macrocell (RMa) for Line-of-Sight and Non Line-of-Sight). In order to perform this analysis, four features are taken into account: Path loss, K Factor, RMS delay spread and RMS angular spread.
  
  \item Thrane et al. \cite{DL_PL} propose deep learning techniques to predict the path loss in the propagation channel at 2.6~GHz. A DNN is able to learn from input as distances, positions and satellite images. The DNN output infers the radio quality parameters, which estimate the path loss of the communication channel.
  
  \item Yang et al. \cite{ML_IVC} classify four scenarios related to vehicular communications, i.e., urban, highways, NLoS channels and tunnels. They use a back-propagation neural network (BPNN) and a feature set formed by four features: power delay spectrum, shadow fading, RMS delay spread and K Factor.
  
\end{itemize}

%%%%%%%%%%%%%%%%%%%%%%%%%%%%%%%%%%%%%%%%%%%%%%%%%%
\begin{figure*}[b]
    \centering
    \includegraphics[width=0.77\textwidth]{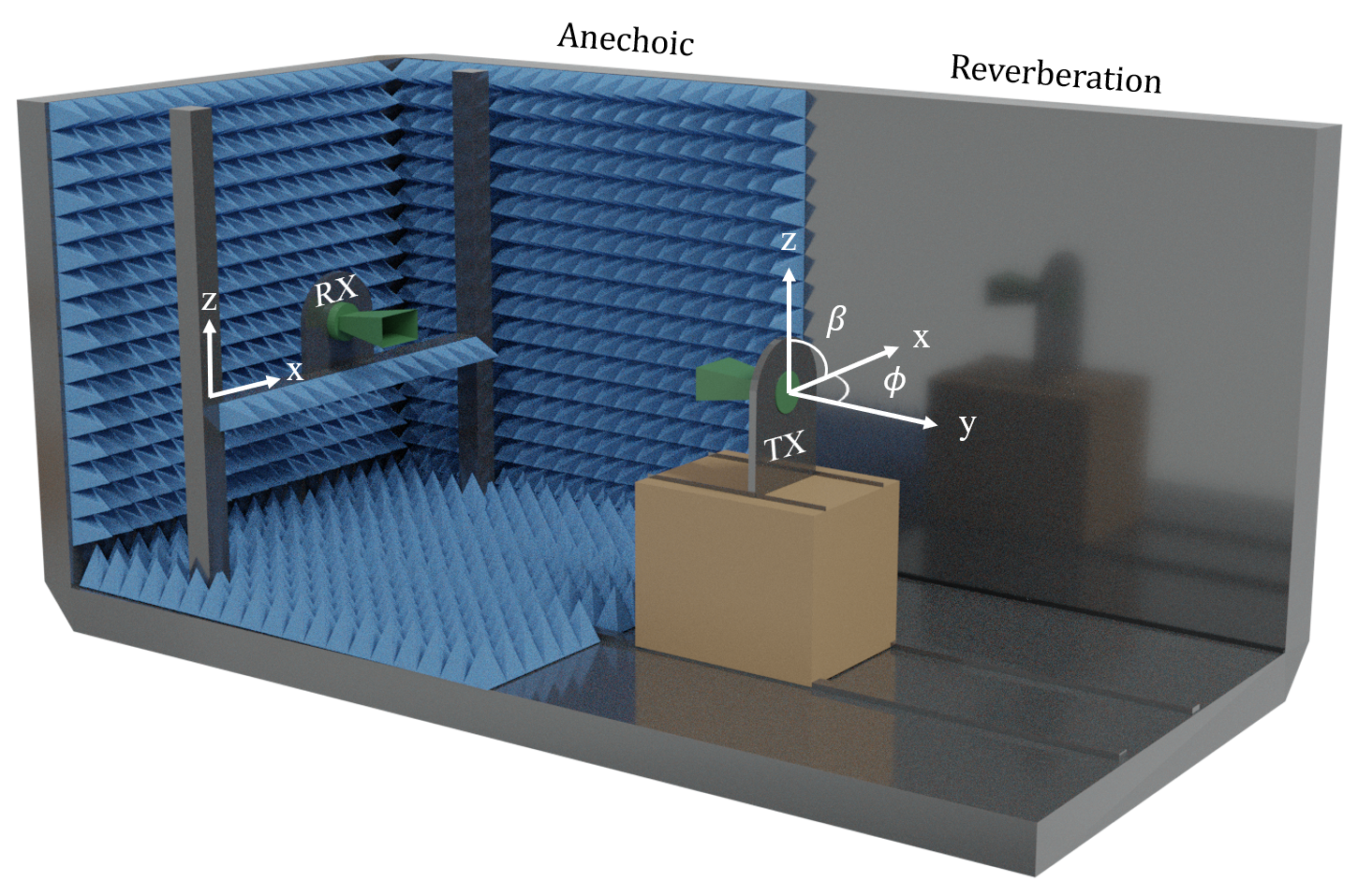}
    \caption{Scheme of the semi-anechoic and semi-reverberation chamber. The scheme shows an anechoic setup, where the transmitter points \blue{at} the receiver in LoS. In a reverberation setup, the transmitter is rotated 180$\degree$ \blue{and it points at} the metallic wall. The azimuth ($\phi$) and roll ($\beta$) angles correspond to the rotation of the transmitter in the XY and XZ planes, respectively. The receiver is free to move in the XZ plane.} 
	\label{esquema_camara}
\end{figure*}
%%%%%%%%%%%%%%%%%%%%%%%%%%%%%%%%%%%%%%%%%%%%%%%%%%%

This work combines communication scenarios with AI and ML techniques to propose a tool for clustering, classification and generation of new 5G communication channels. The main contributions of the work are as follows:

\begin{itemize}

  \item We evaluate the potential application of t-SNE, an unsupervised algorithm for dimensionality reduction and clustering, in the communication field. Although t-SNE technique is well known as a clustering tool, it has not been widely applied in this field. To demonstrate its potential as a powerful and promising tool for the scientific community in this field, the authors cluster five different communication scenarios by using six channel communication features. \blue{The DR allows to perform a rapid visual classification of the scenarios.} These scenarios, of diverse nature, include real wireless communication scenarios and measurements acquired in an anechoic and reverberation chamber.
  
  \item A fitness function that chooses the t-SNE hyperparameters is analyzed. This metric gives those hyperparameters which provide the best visualization of clusters in a two-dimensional plane. The clustering of scenarios is able to show the separation of environments in a very visual way. Moreover, it has been found that t-SNE is able to separate a certain scenario itself into several subclasses, showing its potential as clustering technique.
  
  \item The authors make use of post-processing techniques for scenario modification and generation, studying the effect of a modified scenario in the clustering. As previously stated, the recreation of new scenarios arises as one of the main challenges for future mobile communications. \blue{The application of a DR technique validates the scenario emulation. If an emulated scenario is embedded in the same cluster than a real scenario, it indicates that the DR technique considers both real and emulated scenario as similar.} \blue{These} post-processing techniques, together with classification and clustering algorithms open up the possibility of new scenario recreations.
  
\end{itemize}

%In this work, we evaluate the potential application of t-SNE, an unsupervised algorithm for dimensionality reduction and clustering, in the propagation field. Although t-SNE technique is well known as a clustering tool, it has not been widely applied in the propagation field. To demonstrate its potential as a powerful and promising tool for the scientific community in this field, the authors cluster five different propagation scenarios by using six channel propagation features. The scenarios, of diverse nature, include real wireless communication scenarios and measurements acquired in an anechoic and reverberation chamber. The clustering of scenarios is able to show in a very visual way the separation of environments. Moreover, it has been found that t-SNE is able to separate a certain scenario itself into several subclasses, showing its potential as clustering technique. In addition, the authors make use of post-processing techniques for scenario modification, studying the effect of a modified scenario in the clustering. As previously stated, the recreation of new scenarios arises as one of the main challenges for future mobile communications. This post-processing techniques, together with classification and clustering algorithms open up the possibility of new scenario recreations.

The paper is organized as follows. Section II explains the measurement scenarios that are analyzed throughout the study and the domains in which the communication channels scenarios are presented. Section III presents the clustering technique, t-SNE, for the dimensionality reduction and the input propagation parameters for this algorithm. Section IV analyzes the results obtained from the clustering technique in several communication environments. \blue{Section V provides a comparison among several DR techniques for the analyzed environments. Section VI applies post-processing techniques in order to recreate propagation channel scenarios.} Finally, Section VII summarizes the conclusions extracted in this work.

%%%%%%%%%%%%%%%%%%%%%%%%%%%%%%%%%
\section{\label{sec:Scenarios} Measurement Scenarios}
%%%%%%%%%%%%%%%%%%%%%%%%%%%%%%%%%

This section describes the measurements and environments that are studied and depicted throughout this work. A distinction must be made between two main types of measurements. On the one hand, controlled measurements are acquired in anechoic and reverberation environments\blue{,} where the physical conditions of the propagation channel are managed. On the other hand, real measurements are collected in scenarios that could become part of real communications. Finally, this section explains the two domains (time and frequency) in which the propagation channel is analyzed.

%%%%%%%%%%%%%%%%%%%%%%%%%%%%%%%%%
\subsection{\label{sec:Scenarios_A} Anechoic and Reverberation chamber}
%%%%%%%%%%%%%%%%%%%%%%%%%%%%%%%%%

The first group of channel measurements has been acquired in the facilities of the Smart Wireless Applications and Technologies (SWAT) research group, located at the University of Granada, Spain. The facilities consist of a half anechoic and half reverberation chamber with dimensions 5 $\times$ 3.5 $\times$ 3.5 meters (61.25 m$^3$). Fig. 1 shows a three-dimensional view of the chamber. The semi-anechoic part is fully covered with absorbers \blue{which avoid} any reflections. Thus, a receiver (RX) located inside the anechoic environment \blue{that is} pointed by a transmitter (TX), receives exclusively the electromagnetic wave from the Line-of-Sight (LoS) path. The semi-reverberation part is composed of metallic walls. The presence of the metallic walls provokes that the incident wave is reflected and diffracted into several multipath components (MPCs) that reach the RX at different times. This chamber has been thoroughly described, used and validated in previous works for antenna and propagation measurements \cite{time_gating}-\hspace{1sp}\cite{angel_2}.

%%%%%%%%%%%%%%%%%%%%%%%%%%%%%%%%%%%%%%%%%%%%%%%%%%
\begin{figure*}[b]
    \centering
    \subfigure[]{\includegraphics[width= 0.32\textwidth]{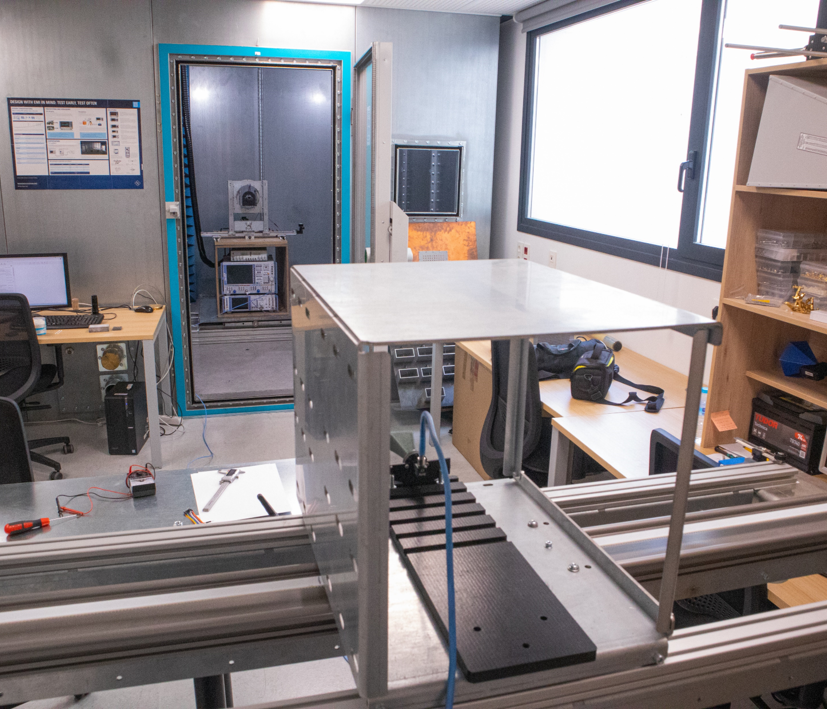}
	}
	\subfigure[]{\includegraphics[width= 0.32\textwidth]{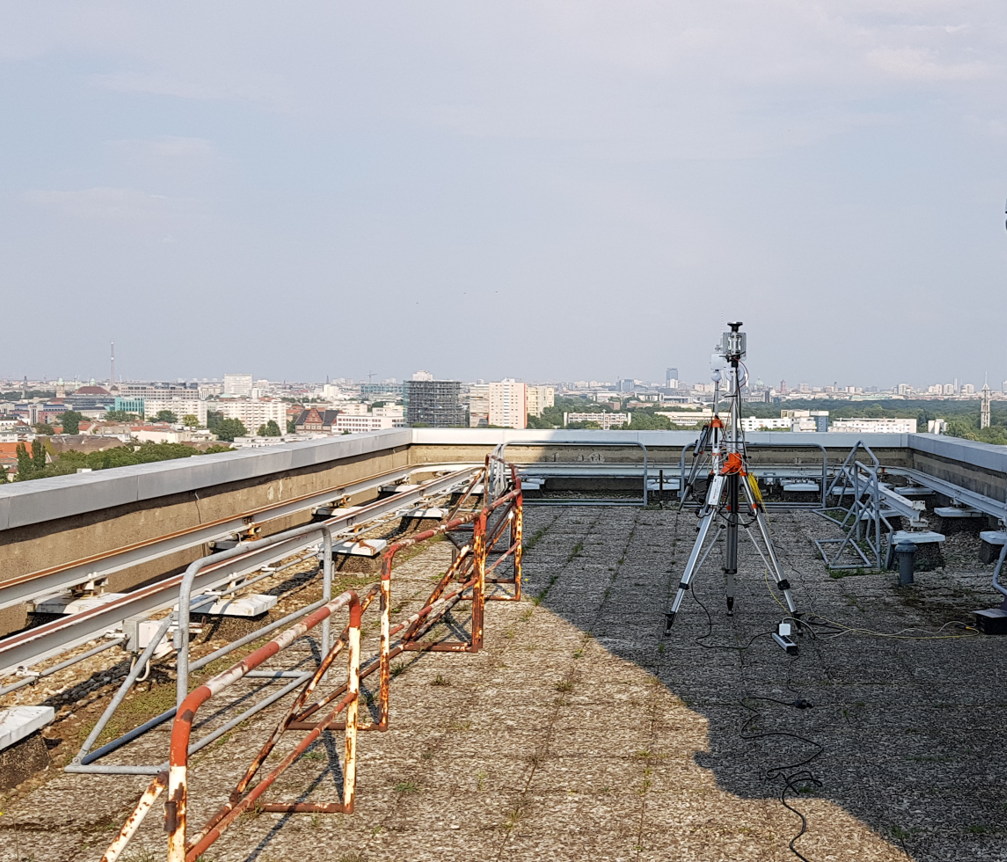}
	}
	\subfigure[]{\includegraphics[width= 0.32\textwidth]{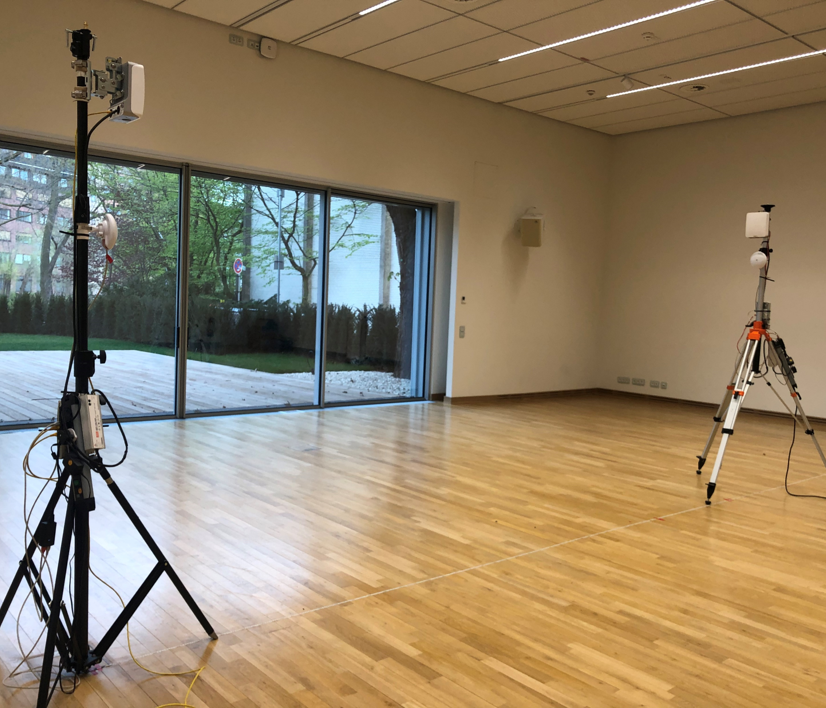}
	}
    \caption{Photographs of the (a) indoor, (b) rooftop and (c) auditorium scenarios that form the dataset. Indoor photograph is taken from the RX perspective. Rooftop \cite{database_roof} and auditorium \cite{database_auditorium} scenarios have been acquired \blue{at} the Fraunhofer-Heinrich-Hertz-Institut.} 
	\label{fotos_escenarios}
\end{figure*}
%%%%%%%%%%%%%%%%%%%%%%%%%%%%%%%%%%%%%%%%%%%%%%%%%%%

In this work, several communication channels are measured for both anechoic and reverberation environments to create a dataset that allows the study of the clustering of scenarios. \blue{For that purpose, the measurement system can move both the receiver and the transmitter. The first one can be shifted in the XZ plane, while the transmitter can be moved in y axis, azimuth ($\phi$) and roll ($\beta$) angles (see Fig. 1).} The receiver, initially aligned with the transmitter, is moved in 4 cm steps in the XZ plane, for a total of 11 positions in each axis, resulting in 121 positions that form a square of dimensions 40 x 40 cm. The transmitter adopts three azimuth positions (-30º, 0º and 30º) in order to modify the pointing angle. In addition, three roll angles (-30º, 0º and 30º) are measured to take into account the polarization effect. Zero degrees in the azimuth angle stand for the LoS path between the pair TX-RX, negative angles for counterclockwise rotation and positive angles for clockwise rotation. Zero degrees in the roll angle imply no depolarization losses between TX and RX. The combination of all possible configurations of the measurement system provides 1089 scenarios for both anechoic and reverberation cases. For the anechoic measurements, the distance between antennas has been set to 160 cm. For the reverberation measurements, the distance between antennas is 600 cm, corresponding to the LoS between the transmitter and the metallic wall plus the return path to the receiver. \blue{Note that TX-RX antenna range is different for the anechoic and reverberation cases. This fact increases the diversity of the measured communication channels due to the different ToA (Time of Arrival). Real communication scenarios consider several ToAs, which have high influence on the channel response. Therefore, this can be a critical parameter in scenario clustering.} These measurements generate the first two environments that will be evaluated: (i) anechoic and (ii) reverberation.

The acquisition for environments (i) and (ii) is performed with a Vectorial Network Analyzer (VNA Rohde \& Schwarz ZVA67), which measures the scattering parameters in the propagation channel and can operate up to 67 GHz. The chosen antennas for the acquisition are standardized gain horns fed with a WR-34 waveguide-to-coaxial transition (Flann Kband antenna Model: \#21240-20) for both transmitter and receiver. Radiating patterns of these elements can be found in \cite{time_gating}. Prior to the acquisition process, a TOSM \blue{(Through – Open – Short - Match)} calibration is performed to eliminate the effect of the coaxial cable. Therefore the scattering parameters are measured from the aperture of the radiating elements. The transmit power is set to 10 dBm in the VNA. The frequency range for these scenarios goes from 24.25 GHz to 27.5 GHz, for a total of 651 frequency samples. This provides 5 MHz frequency separation. This band, also called 3GPP n258 \cite{3GPP_TS}, is considered fundamental for the deployment of 5G and millimeter wave communications in the European Union.

%%%%%%%%%%%%%%%%%%%%%%%%%%%%%%%%%
\subsection{\label{sec:Scenarios_B} Indoor and Outdoor scenarios}
%%%%%%%%%%%%%%%%%%%%%%%%%%%%%%%%%

In addition to the anechoic and reverberation scenarios, three \blue{additional} scenarios that could be part of real communication environments will be evaluated. 

\begin{enumerate}[topsep=1pt,itemsep=1ex,partopsep=1ex,parsep=1ex,label=(\roman*)]
  \setcounter{enumi}{2}

  \item The third one is an indoor scenario located in the same facilities than \blue{those from} Section II.A. In this case, the transmitter points at a furnished laboratory through the chamber door [see Fig. 2(a)]. The receiver, placed in a measurement system similar to the one in the previous section, has mobility in XZ plane. Therefore, we consider 121 positions in a square of dimension 40 $\times$ 40 cm. The transmitter moves in the azimuth ($-30\degree$, $-15\degree$ and $0\degree$) and roll angles ($-30\degree$, $0\degree$ and $30\degree$) for a total combination of 1089 configurations. When $\phi = -30\degree$, TX points through the window on the right side in Fig. 2(a). When $\phi = -15\degree$, TX points towards the chamber door frame. Finally, the case $\phi = 0\degree$ describes the angle where TX points directly at RX through the chamber door in LoS. The measurement configuration, i.e., radiating elements, calibration \blue{process}, transmitted power, frequency range and frequency samples, is similar to the one shown in environments (i) and (ii).
  
  \item The fourth environment consists of a dataset from \blue{a rooftop outdoor scenario located at} the Fraunhofer-Heinrich-Hertz-Institut in Berlin, Germany \cite{database_roof} [see Fig. 2(b)]. Two antennas are aligned in LoS with a separation of 800 cm. Both TX and RX are steered in azimuth angle from -45º to 45º with 64 possible positions for each antenna. Moreover, the dataset is increased by acquiring measurements placing the TX and RX misaligned between them and also switched. In order to keep a balance in the number of measurements of each type, 1089 of the total number of measurements are selected. The signal acquisition technique is based on real-time sampling, where the sampling rate is 3.52 GHz. The transmitted signal is centered in the millimeter wave band at 60.48 GHz.
  
  \item The fifth environment comes from a dataset measured in an auditorium at the Fraunhofer-Heinrich-Hertz-Institut \cite{database_auditorium}. Similar to the previous scenario, two antennas are aligned in LoS and the distance between them is chosen to be 700 cm and 800 cm. TX and RX can point in the vertical axis from -45º to 45º degrees with 64 positions. 1089 measurements from this dataset are selected. The acquisition properties and operation frequencies are similar to those of the rooftop scenario.
  
\end{enumerate}

The (iv) rooftop and (v) auditorium communication scenarios have been acquired at the Fraunhofer-Heinrich-Hertz-Institut in Berlin, Germany. The datasets have been made public through the NextG Channel Model Alliance. Further information about the dataset, i.e., the acquisition \blue{process}, the channel sounder architecture and the scenario physical parameters for environments (iv) and (v), is thoroughly detailed in \cite{database_roof},\cite{database_auditorium}.

%%%%%%%%%%%%%%%%%%%%%%%%%%%%%%%%%
\subsection{\label{sec:Scenarios_C} Propagation channel in temporal and frequency domain}
%%%%%%%%%%%%%%%%%%%%%%%%%%%%%%%%%

In order to perform the clustering that will be shown in Section IV, we need to extract several properties from the communication channels that allow the differentiation of the five proposed scenarios. For that purpose, the propagation channels can be analyzed in two domains, the temporal domain through the Channel Impulse Response (CIR) and the frequency domain through the Channel Frequency Response (CFR).

Traditionally, CIR is \blue{modeled} as follows \cite{propagation_channel}:

\begin{equation}
    \blue{h(t_n)=\sum_{k=0}^{K-1} \alpha_{k} \delta\left(t_n-\tau_{k}\right)}
\end{equation}

\noindent where \blue{$t_n$ refers to the $n$-th time sample}, $k$ is the index of considered sample, $K$ is the number of samples in the time domain,  $\delta(\cdot)$ stands for the Dirac delta function, $\tau_{k}$ shows the delay of the arrival in the propagation channel between the pair TX-RX and $\alpha_{k}$ denotes the complex amplitude which takes into account the attenuation and phase change due to the physical phenomena (reflection, scattering, refraction, diffraction) that occur during signal propagation.

The CIR has its equivalent in the frequency domain through the CFR:

\begin{equation}
H(f)=\sum_{k=0}^{K-1} b_{k} e^{-j 2 \pi f \tau_{k}}
\end{equation}

\noindent where $b_{k}$ stands for the complex amplitude of the CFR, and $e^{-j 2 \pi f \tau_{k}}$ denotes the complex exponential that depends on the frequency and the time of arrival.

Once a measurement in one of the two domains has been acquired, Discrete Fourier Transform (DFT) and Inverse Discrete Fourier Transform (IDFT) \blue{allow the calculation of} the CFR from the CIR [eq. (3)] and vice versa [eq. (4)].

\begin{equation}
H(f)=\sum_{n=0}^{\frac{t_{\max }}{T}} h\left(t_{n}\right) e^{-j 2 \pi f t_{l}}
\end{equation}

\begin{equation}
h(t)=\sum_{l=0}^{\frac{B}{\Delta f}} H\left(f_{l}\right) e^{j 2 \pi f_{l} t}
\end{equation}

In eq. (3), $t_{\max}$ is the time of the last sample acquired and T is the sampling period. In eq. (4), $B$ stands for the total bandwidth of the measurement band and $\Delta f$ is the frequency step. Therefore, $t_{\max}/{T}$ and $B/\Delta f$ factors are the total number of points acquired in the time and frequency domain, respectively.

As previously explained, environments (i), (ii) and (iii) are analyzed through the scattering parameters acquired in the VNA. Therefore, these environments follow the notation of the eq.~(2). On the other hand, environments (iv) and (v) are acquired through a channel sounder in the time domain. Consequently, the acquired data can be expressed as eq.~(1). Since all environments need to be expressed in both domains to extract the propagation features that will be shown in Section~III, we apply eq.~(4) to the environments (i), (ii) and (iii), and eq.~(3) to the environments (iv) and (v).

%%%%%%%%%%%%%%%%%%%%%%%%%%%%%%%%%
\section{\label{sec:AI} AI for propagation measurements}
%%%%%%%%%%%%%%%%%%%%%%%%%%%%%%%%%

As stated in Section I, Artificial Intelligence is becoming one \blue{of} the main topics in several research fields \cite{tSNE_recognition}, \cite{tSNE_diagnosis}. In wireless communications, AI emerges as a potential tool for scenario recreation and generation in future mobile communications. This section shows the clustering algorithm that is implemented, the hyperparameters involved in its performance and the channel features that are considered as the input of this algorithm.

%AI can learn to classify and cluster several types of scenarios.

%%%%%%%%%%%%%%%%%%%%%%%%%%%%%%%%%
\subsection{\label{sec:AI_A}An overview of t-SNE}
%%%%%%%%%%%%%%%%%%%%%%%%%%%%%%%%%

The t-distributed Stochastic Neighbor Embedding \cite{tSNE}, \cite{tSNE_accelerating}-\hspace{1sp}\cite{tSNE_progressive} is an unsupervised learning algorithm that reduces the dimensionality of a high-dimensional dataset $\lbrace \mathbf{x}_1,\mathbf{x}_2,\ldots,\mathbf{x}_N \rbrace$ into a low-dimensional (2D) dataset $\lbrace \mathbf{y}_1,\mathbf{y}_2,\ldots,\mathbf{y}_N \rbrace$. The high-dimensional space is defined by the matrix $\mathbf{X} \in \mathbb{R}^{N \times F}$, where $N$ is the number of communication channel measurements and $F$ is the number of communication channel parameters contained in the high-dimensional space. Each row of $\mathbf{X}$ contains the channel parameters of the $i$-th observation; namely,  $\mathbf{x}_i$ is a row vector that can be defined as

\begin{equation}
\mathbf{x}_i = [K_{i}, \tau_{RMS,i}, \tau_{\text {mean,}i}, \tau_{\text {var,}i}, \overline{PL}_{i}, \bar{\eta}_i]
\end{equation}

\noindent These communication channel parameters will be identified later on Section III.B. Similarly, the low-dimensional space is defined by the matrix $\mathbf{Y} \in \mathbb{R}^{N \times 2}$. The dimension of this space is set to two due to the ease of visualization in a two-dimensional plane.

t-SNE arises as an improvement to the Stochastic Neighbor Embedding (SNE), where the distances between datapoints in the high-dimensional dataset are \blue{modeled} as joint probability distributions $p_{ij}$, known as the similarity between the datapoints. Therefore, the similarity for every pair $\mathbf{x}_{i}$ and $\mathbf{x}_{j}$ is computed as the probability $p_{ij}$, calculated as follows:

\begin{equation}
p_{j \mid i}=\frac{\exp \left(-\left\|\mathbf{x}_{i}-\mathbf{x}_{j}\right\|^{2} / 2 \sigma_{i}^{2}\right)}{\sum_{k \neq i} \exp \left(-\left\|\mathbf{x}_{i}-\mathbf{x}_{k}\right\|^{2} / 2 \sigma_{i}^{2}\right)}
\end{equation}

\begin{equation}
p_{i j}=\frac{p_{j \mid i}+p_{i \mid j}}{2 N}
\end{equation}

\noindent where $\sigma_{i}^{2}$ is the variance of a Gaussian probability density function (PDF) for the normal distribution centered in the datapoint $\mathbf{x}_{i}$, $p_{j \mid i}$ and $p_{i \mid j}$ are the conditional probabilities between the datapoints $\mathbf{x}_{i}$ and $\mathbf{x}_{j}$, and $\left\| \cdot \right\|$ stands for the \blue{E}uclidean norm.

For the low-dimensional dataset, t-SNE uses a Student’s t-distribution in order to define a joint probability denoted by $q_{ij}$ and calculated as:

\begin{equation}
q_{i j}=\frac{\left(1+\left\|\mathbf{y}_{i}-\mathbf{y}_{j}\right\|^{2}\right)^{-1}}{\sum_{k \neq l}\left(1+\left\|\mathbf{y}_{k}-\mathbf{y}_{l}\right\|^{2}\right)^{-1}}
\end{equation}

Once both joint probabilities $p_{ij}$ and $q_{ij}$ have been defined, t-SNE minimizes the Kullback-Leibler divergence [KL($\cdot$)] between joint probability distributions in the high-dimensional ($\mathbf{P}$) and low-dimensional ($\mathbf{Q}$) spaces. Intuitively, large $p_{ij}$ values [eq. (7)] indicate that $\mathbf{x}_{i}$ and $\mathbf{x}_{j}$ are closer in the high-dimensional space. In the same way, large $q_{ij}$ values [eq. (8)] indicate that $\mathbf{y}_{i}$ and $\mathbf{y}_{j}$ are closer in the low-dimensional space. If the dimensionality reduction performs a proper mapping, the probability distributions $\mathbf{P}$ and $\mathbf{Q}$ should resemble each other. In order to quantify the quality of the dimensionality reduction, a cost function is defined as \cite{tSNE}:

\begin{equation}
C=\operatorname{KL}(\mathbf{P} \| \mathbf{Q})=\sum_{i} \sum_{j} p_{i j} \log _{2} \frac{p_{i j}}{q_{i j}}
\end{equation}

The minimization of this function indicates an accurate mapping between the high-dimensional (probability distribution $\mathbf{P}$) and low-dimensional (probability distribution $\mathbf{Q}$) spaces. The minimization is made by using a gradient descend technique, where the gradient has the following form \cite[Appx.~A]{tSNE}:

\begin{equation}
\frac{\delta C}{\delta \mathbf{y}_{i}}=4 \sum_{j}\left(p_{i j}-q_{i j}\right)\left(\mathbf{y}_{i}-\mathbf{y}_{j}\right)\left(1+\left\|\mathbf{y}_{i}-\mathbf{y}_{j}\right\|^{2}\right)^{-1}
\end{equation}

Through several iterations of the algorithm, the dimensionality reduction of the high-dimensional dataset $\lbrace \mathbf{x}_1,\mathbf{x}_2,\ldots,\mathbf{x}_N \rbrace$ into a low-dimensional dataset $\lbrace \mathbf{y}_1,\mathbf{y}_2,\ldots,\mathbf{y}_N \rbrace$ is improved. A detailed explanation of the whole technique can be found in \cite{tSNE}.

One of the key features of this dimensionality reduction technique is its flexibility given a set of configuration hyperparameters. Previous works made by the authors \cite{tSNE_EuCAP} have shown the relevance of a correct choice of hyperparameters. Particularly, we detected three hyperparameters which are crucial in order to find a proper dimensionality reduction. These parameters are\blue{:} the type of distance between datapoints, the perplexity, and the learning rate. A brief explanation, as well as its relation to the previous formulation, is given below.

\begin{enumerate}[topsep=1pt,itemsep=1ex,partopsep=1ex,parsep=1ex,label=$\bullet$]
  \item Distance: Previous equations [eqs. (6), (8), (10)] consider the Euclidean distance between datapoints. When the variance between different features in the high-dimensional dataset $\lbrace \mathbf{x}_1,\mathbf{x}_2,\ldots,\mathbf{x}_N \rbrace$ is in a different range (see Subsection III.B), the Euclidean distance does not give the same importance to all the variables. In order to avoid this fact, the Mahalanobis distance takes into account the covariance matrix $\mathbf{\xi}$. Therefore, this distance is used throughout the work. It is defined as:
  
  \begin{equation}
    d_{m}(\mathbf{x}_i, \mathbf{x}_j)=\sqrt{(\mathbf{x}_i-\mathbf{x}_j) {\xi}^{-1}(\mathbf{x}_i-\mathbf{x}_j)^{T}}
  \end{equation}
  
  \noindent \blue{where $\mathbf{\xi}$ is the  sample covariance of matrix $\mathbf{X}$ with dimensions $F \times F$.}
  
  \item Perplexity: In eq. (6), the conditional probability depends on the variance of a Gaussian PDF $\sigma_{i}$. This value is chosen such that the joint probability for all the datapoints is fixed to a certain perplexity, which is defined as:
  
  \begin{equation}
    \operatorname{Perp}_i=2^{\psi_i}
  \end{equation}
  
  \begin{equation}
    \psi_i=-\sum_{j} p_{j \mid i} \log _{2}\left(p_{j \mid i}\right)
  \end{equation}
  
  where $\psi_{i}$ stands for the Shannon entropy. Since there is not a single perplexity value that provides the optimum performance, Section IV shows a search for a proper value of this parameter in terms of visualization for our datasets.
  
  \item Learning rate: This value is related to the convergence of the algorithm through several iterations. For each iteration, the learning rate $\rho$ multiplies the gradient shown in eq. (10) \cite{tSNE}. On the one hand, if this value is too small, the gradient descent could be slow. On the other hand, if this value is too high the gradient descent might not converge correctly. In Section IV, a proper value of this parameter is discussed for the presented datasets.
  
\end{enumerate}

%%%%%%%%%%%%%%%%%%%%%%%%%%%%%%%%%
\subsection{\label{sec:AI_B}Communication channel parameters for t-SNE}
%%%%%%%%%%%%%%%%%%%%%%%%%%%%%%%%%

In this paper, five datasets have been shown in Section~II, containing 1089 communication channel measurements each one, for a total of 5445. This leads to a six-dimensional dataset composed by 5445 observations $\lbrace \mathbf{x}_1,\mathbf{x}_2,\ldots,\mathbf{x}_{5445} \rbrace$. Each measurement is characterized by six communication channel parameters:

\begin{enumerate}[topsep=1pt,itemsep=1ex,partopsep=1ex,parsep=1ex,label=$\bullet$]

  \item $K$ Factor: It represents the relation between the dominant multipath component, usually the LoS, and the rest of multipath components in the time domain. It is approximated as the ratio between the maximum power component $| h_{i}(t_{n_\mathrm{max}})|^{2}_{\max }$ and the \blue{sum power of the remaining taps in the power delay profile (PDP).}
  
  \begin{equation}
    K_{i}=10 \log _{10}\left(\frac{\left|h_{i}(t_{n_\mathrm{max}})\right|^{2}_{\max }}{\sum_{n=0, 
     \\ n \neq n_\mathrm{max}} \left|h_{i}(t_n)\right|^{2}}\right)
  \end{equation}
  
  \item $\tau_{mean}$: The mean delay is defined as the first moment of the PDP \cite{recommendation_ITU}. It shows the average delay of the power measured in the communication channel.
  
  \begin{equation}
    \tau_{\text {mean,}i}=\frac{\sum_{n=0} t_n \cdot\left|h_{i}(t_n)\right|^{2}}{\sum_{n=0}\left|h_{i}(t_n)\right|^{2}}
  \end{equation}
  
  \item $\tau_{var}$: The variance delay is obtained as the second moment of the PDP and depicts how fast the PDP power varies in short time intervals.
  
  \begin{equation}
    \tau_{\text {var},i}=\frac{\sum_{n=0} {t_{n}}^{2} \cdot\left|h_{i}(t_n)\right|^{2}}{\sum_{n=0}\left|h_{i}(t_n)\right|^{2}}
  \end{equation}
  
  \item $\tau_{RMS}$: The root-mean-square delay spread is calculated as the second central moment root square of the PDP \cite{recommendation_ITU} and presents the power deviation of the communication channel. Large values indicate that the power is divided into several MPCs. On the other hand, small values are representative of scenarios where the power is concentrated on the main MPC, typically the Line-of-Sight. In other words, this feature describes the power dispersion in the time domain of the communication channel.
  
  \begin{equation}
    \tau_{RMS,i}=\sqrt{\frac{\sum_{n=0}\left(t_n-\tau_{\text {mean,}i}\right)^{2} \cdot\left|h_{i}(t_n)\right|^{2}}{\sum_{n=0}\left|h_{i}(t_n)\right|^{2}}}
  \end{equation}
  
  \item Path Loss (PL): This parameter shows the attenuation between the pair TX-RX due to propagation losses through the communication channel. For this study, this value includes the contribution of the antenna gain from the TX and RX antennas. This value is calculated as the absolute value of the channel frequency response averaged over the complete domain of the CFR, i.e., averaged over all frequencies ($\overline{PL}_i$).
  
    \begin{equation}
    PL_{i}(f)=20\log _{10}\left(|\blue{H_i(f)}|\right)
  \end{equation}
  
  \item Spectral Efficiency ($\eta$): The spectral efficiency is defined as the information rate that can be sent through a communication system, i.e., a propagation channel. It is measured in bps/Hz and calculated as follows \cite{fundamentals_massive}:
  
%  \begin{equation}
%    \eta_{i}=\log _{2}\left| \det \Big(I_{RX}+\frac{SNR}{T X} H_{i}(f)^{H} H_{i}(f)\Big)\right|
%  \end{equation}

  \begin{equation}
    \eta_{i}(f)=\log _{2}\left( 1 + SNR \cdot |H_i(f)|^2\right)
  \end{equation}
  
\blue{where SNR stands for the Signal-to-Noise Ratio.} Then, $\eta_{i}(f)$ is averaged over the whole frequency range of the measurement. The average spectral efficiency is noted as~$\bar{\eta}_{i}$.
  
  \end{enumerate}

Notice that the former expressions [eqs. (14)-(19)] stand for the $i$-th observation. Once every communication channel feature has been calculated, they form the six-dimensional space $\mathbf{x}_i$ for the $i$-th observation.

%Finally, the high-dimensional dataset with 5445 measurements (rows) and 6 propagation channel parameters (columns) works as the t-SNE input.

%%%%%%%%%%%%%%%%%%%%%%%%%%%%%%%%%
\section{\label{sec:Results} Clustering Results}
%%%%%%%%%%%%%%%%%%%%%%%%%%%%%%%%%

Once the theoretical basis and t-SNE have been explained, this section presents and discusses the results obtained for several clustering of communication scenarios, where each subsection shows the results for a specific data subset. Good scenario separation using clustering techniques involves the unambiguous identification of communication environments. This fact can be crucial in the classification and generation of future communication scenarios, such as UAV-to-UAV, S2S or V2V systems.

As demonstrated in \cite{tSNE_EuCAP}, the Mahalanobis distance is the distance metric that provides the best dimensionality reduction. However, clusters are also altered by the learning rate and perplexity. For these two hyperparameters, the choice of a proper value in terms of visualization is not trivial. In order to solve this fact, we define a fitness function $F$, which measures the ratio between the inter class distance and intra class distance. Intra class distance is described as the sum of distances in the low-dimensional space between datapoints belonging to the same class. On the other hand, inter class distance is defined as the sum of distances in the low-dimensional space between datapoints from different classes. If we use this fitness function, we are able to find those configuration parameters that provide the largest separation between classes. By maximizing this metric, the visualization of the figures is easier for the reader due to the formation of clear clusters. Mathematically, this fitness function can be defined as:

%\red{\textit{En esta afirmación te estás tirando piedras sobre tu propio tejado. Habría que suavizarla.} Although it does not provide physical information to draw conclusions from the distances between classes, the reason for using this metric is to make it easier for the reader the visualization of the figures.}

\begin{equation}
F^{(\mathscr{X})}=\frac{\sum_{i \in \mathscr{X}} \sum_{\forall j \notin \mathscr{X}} d_{e}(i, j)}{\sum_{i \in \mathscr{X}} \sum_{k \in \mathscr{X}} d_{e}(i, k)}
\end{equation}

\begin{equation}
F=\frac{1}{\left(N_{c}-1\right)} \sum_{\mathscr{X}=1}^{N_{c}} F^{(\mathscr{X})}
\end{equation}

\noindent  where $d_{e}$ stands for the Euclidean distance, $\mathscr{X}$ is the considered class, $i, j, k$ are indexes represented by natural numbers, and $N_{c} > 1$ is the total number of classes.

For a better visualization of the results, it is intended that the inter class distance becomes as large as possible and the intra class distance becomes as small as possible. Therefore, the main idea is to find those parameters that maximize $F$. As an example of this fitness function, Fig. 3 shows results for the dataset composed by the anechoic, reverberation and indoor datasets for values of learning rate and perplexity \blue{in the range [1, 750] $\in \mathbb{N}$.}

%%%%%%%%%%%%%%%%%%%%%%%%%%%%%%%%%%%%%%%%%%%%%%%
\begin{figure}[t]
	\centering
	\includegraphics[width= 1\columnwidth]{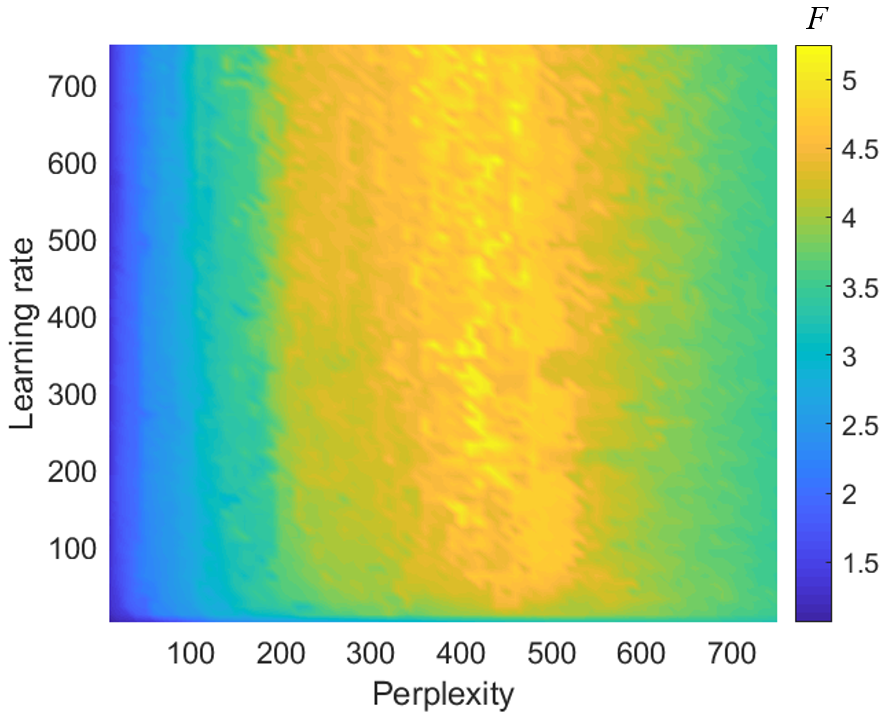}
	\caption{Fitness function for the subset composed by the anechoic, reverberation and indoor datasets (see Section IV.A).} 
	\label{barrido_F}
\end{figure}
%%%%%%%%%%%%%%%%%%%%%%%%%%%%%%%%%%%%%%%%%%%%%%%

Two main conclusions can be drawn from Fig. 3. First, the algorithm tends to converge for learning rate above 15. Below this value, the gradient descent in eq. (10) is too slow and the algorithm is not able to find a good solution. Second, when the learning rate is in a good range, the critical configuration parameter is the perplexity. Too small or too large perplexity values do not find an optimal solution since small values do not properly group the clusters and large values tend to group inter classes, which minimizes the inter class distance. Therefore, the optimal solution lies somewhere in between. In this case, the maximum is found for perplexity values around 420. Specifically, the fitness function maximum is $F = 5.23$ for learning rate and perplexity values equal to 600 and 440. This value means that, in average, the distance between datapoints from different classes is 5.23 times higher than distances between datapoints from the same class. Due to this fact, the visualization of the clusters in the low-dimensional space is straightforward. \blue{These t-SNE hyperparameters (learning rate and perplexity) are detailed on the caption of subsequent figures throughout this Section.}

In addition, a radar chart (see Fig. 10) with the numerical results of the mean and standard deviation of the communication channel properties for each possible scenario is shown at the end of this section. The channel features have a direct impact on the KPIs, i.e. they directly influence the behavior of the communication channel. As an example, the spectral efficiency is related to the network data rate performance. These results are discussed throughout this section to provide a rational explanation for the clustering results.

%%%%%%%%%%%%%%%%%%%%%%%%%%%%%%%%%%%%%%%%%%%%%%%
\begin{figure}[b]
	\centering
	\includegraphics[width= 0.95\columnwidth]{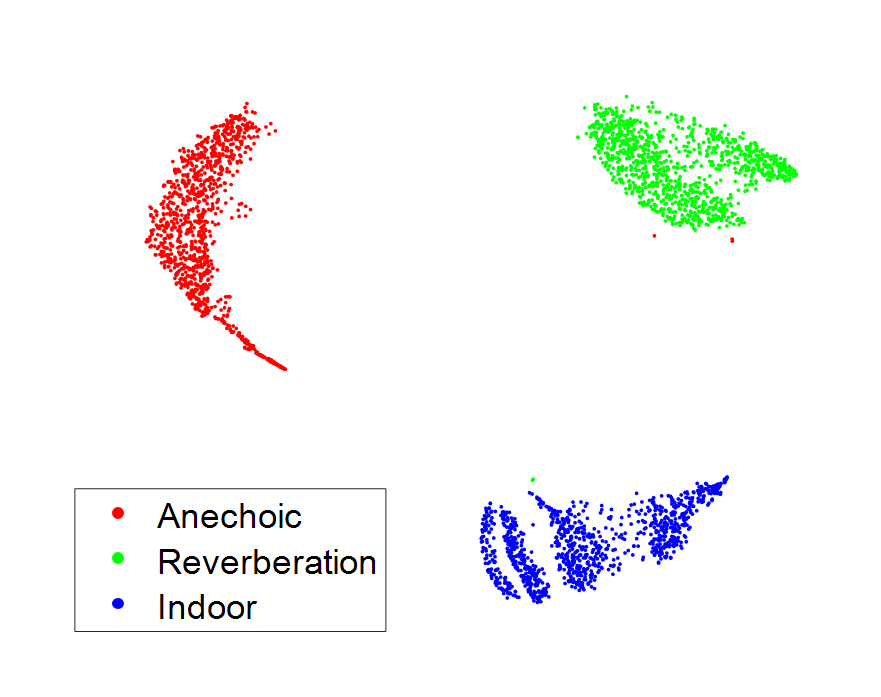}
	\caption{Clustering of anechoic, reverberation and indoor scenarios. Learning rate and perplexity are fixed to 600 and 440.} 
	\label{cluster_tres}
\end{figure}
%%%%%%%%%%%%%%%%%%%%%%%%%%%%%%%%%%%%%%%%%%%%%%%

%%%%%%%%%%%%%%%%%%%%%%%%%%%%%%%%%
\subsection{\label{sec:Results_A}Anechoic, Reverberation and Indoor environments}
%%%%%%%%%%%%%%%%%%%%%%%%%%%%%%%%%

%%%%%%%%%%%%%%%%%%%%%%%%%%%%%%%%%%%%%%%%%%%%%%%
\begin{figure}[t]
	\centering
	\includegraphics[width= 1\columnwidth]{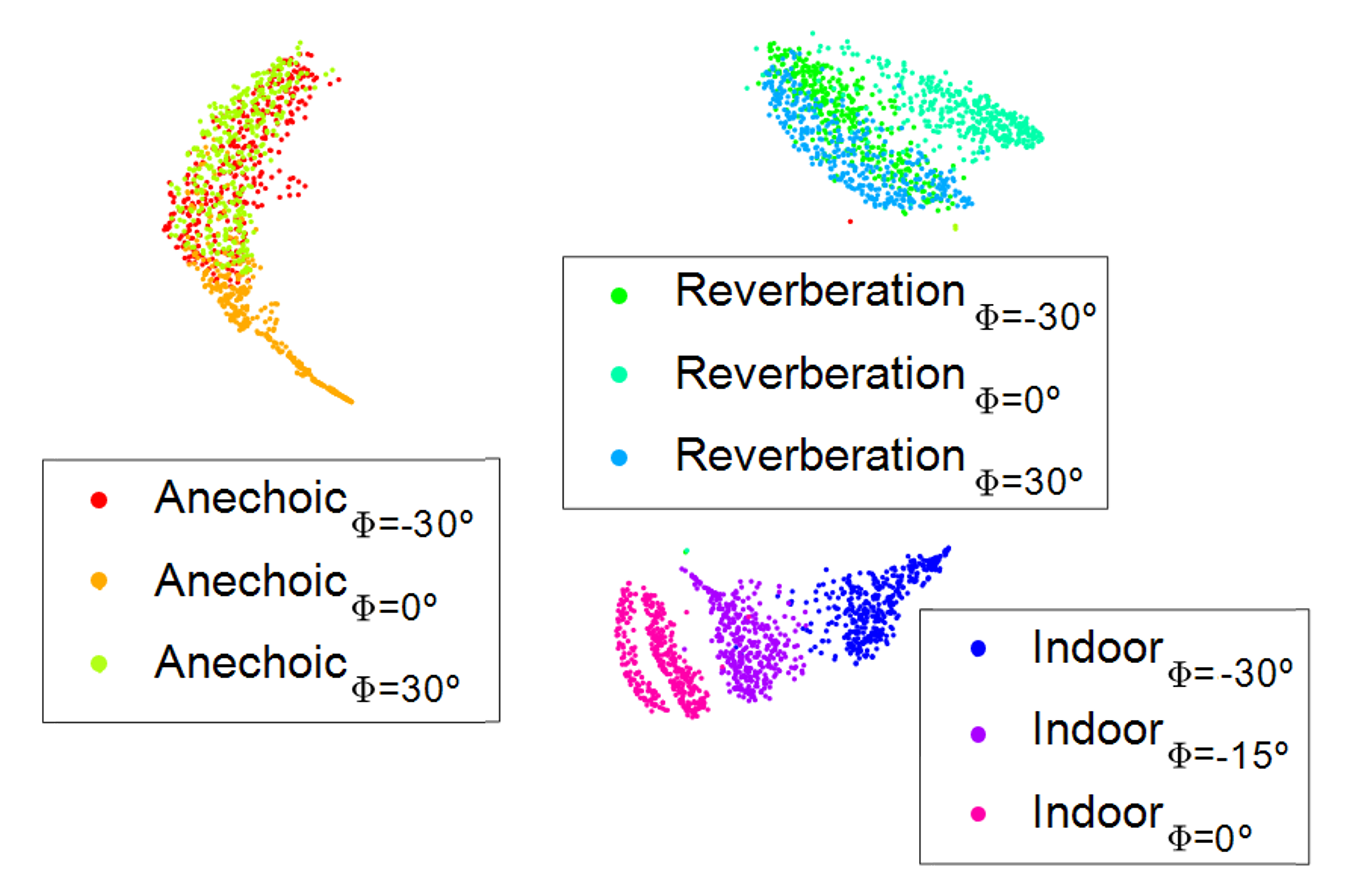}
	\caption{Clustering of anechoic, reverberation and indoor scenarios for several values of the azimuth angle. Learning rate and perplexity are fixed to 600 and 440.} 
	\label{cluster_nueve}
\end{figure}
%%%%%%%%%%%%%%%%%%%%%%%%%%%%%%%%%%%%%%%%%%%%%%%

%%%%%%%%%%%%%%%%%%%%%%%%%%%%%%%%%%%%%%%%%%%%%%%
\begin{figure}[b]
	\centering
	\includegraphics[width= 0.95\columnwidth]{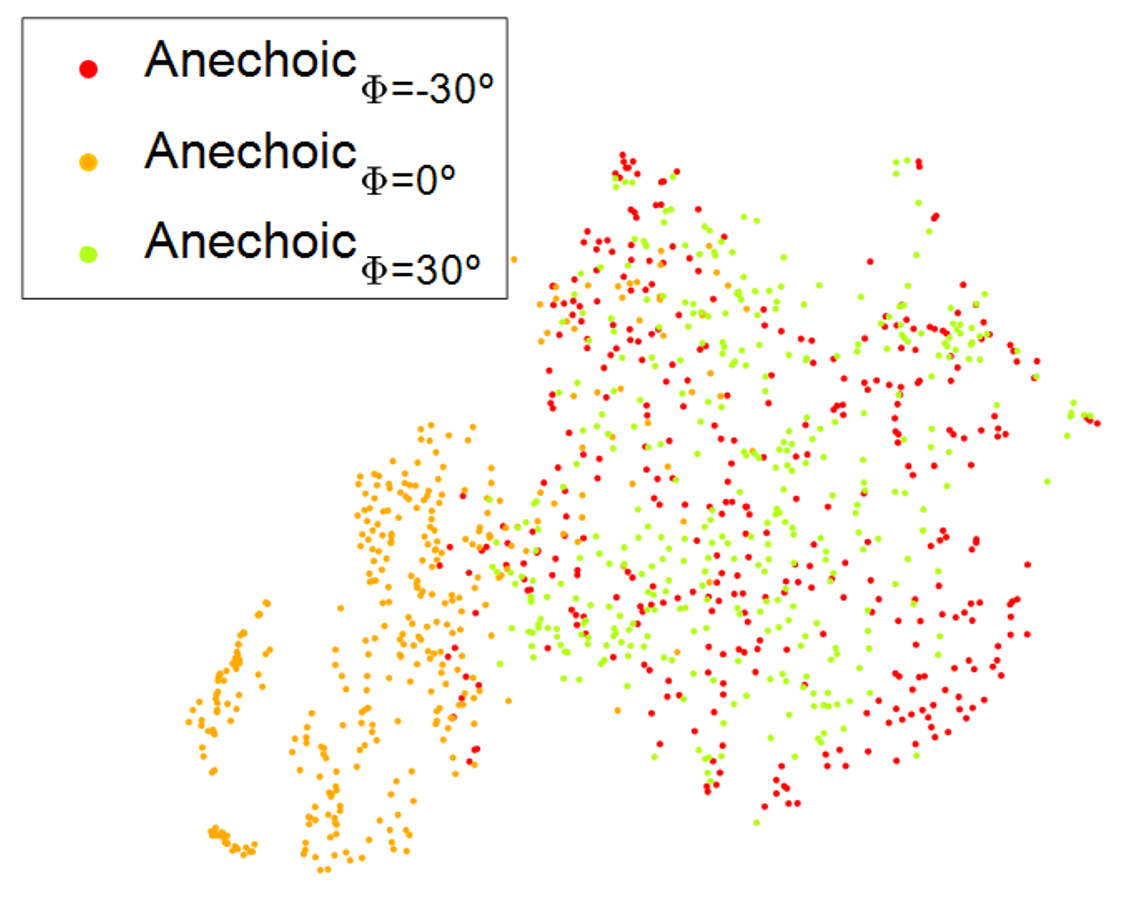}
	\caption{Clustering of the anechoic scenario for three azimuth angles ($-30\degree$, $0\degree$ and $30\degree$). Learning rate and perplexity are fixed to 650 and 240.} 
	\label{cluster_anecoico}
\end{figure}
%%%%%%%%%%%%%%%%%%%%%%%%%%%%%%%%%%%%%%%%%%%%%%%

As a proof of concept, the first clustering includes the anechoic, reverberation and indoor scenarios. These scenarios are characterized by different propagation conditions so that we expect a good separation between clusters. The anechoic environment is typically identified by an unique MPC since the reflections are attenuated by the absorbers. In the reverberation scenario, several MPCs following an exponential decay in the time domain are expected. These MPCs reach \blue{the} RX at discrete times due to the presence of absorbers in the semi-anechoic part of the chamber. Finally, the indoor scenario presents several MPCs following a continuous exponential decay due to the presence of furniture in the room. The clustering presented in Fig. 4 shows an excellent separation in the low-dimensional space when t-SNE is applied for the communication channel parameters previously explained. \blue{Note that a cluster is defined as the two-dimensional area where the datapoint density is high, thus generating a grouping of observations expected to have some common behavior.} Looking at the radar chart (see Fig. 10), this separation is obtained due \blue{to} the following reasons. $\tau_{mean}$ is significantly lower in the anechoic case due to the proximity between antennas (160 cm), compared to the reverberation and indoor cases (600 cm), where multiple reflections \blue{enlarge this distance even further}. The K Factor is substantially larger in the anechoic scenario due to LoS previously stated\blue{.} However, in the reverberation and indoor scenarios, this number tends to be below 0 dB, meaning that the signal is more \blue{spread} across multiple MPCs. $\tau_{RMS}$ and $\tau_{var}$ also play a key role since they are small for the anechoic and reverberation scenarios. This fact can be explained due to the presence of absorbers inside the chamber, obtaining a smoothed PDP that reduces these two values. \blue{In the anechoic case, there is one dominant MPC since the NLoS paths are attenuated by the absorbers. In the reverberation case, several MPCs impinge the RX. However, these MPCs have short delay offsets since large delay offsets MPCs are attenuated by the absorbers. Therefore,} the absence of these absorbers in the indoor scenario induces higher $\tau_{RMS}$ and $\tau_{var}$.

Inspecting Fig. 4, it can be observed that classes present a tendency towards internal separation. Actually, the azimuth angle can be considered as a discriminating parameter, dividing the main class into subclasses. This change leads to the results shown in Fig. 5. Indoor scenario can be separated into three subscenarios depending on the angle, and anechoic and reverberation scenarios tend to separate the central azimuth angle $\phi = 0\degree$ from the others two. Since clear differences are noticeable within the clusters themselves, Section IV.B goes one step beyond and analyzes these differences.

%%%%%%%%%%%%%%%%%%%%%%%%%%%%%%%%%%%%%%%%%%%%%%%
\begin{figure}[b]
	\centering
	\includegraphics[width= 1\columnwidth]{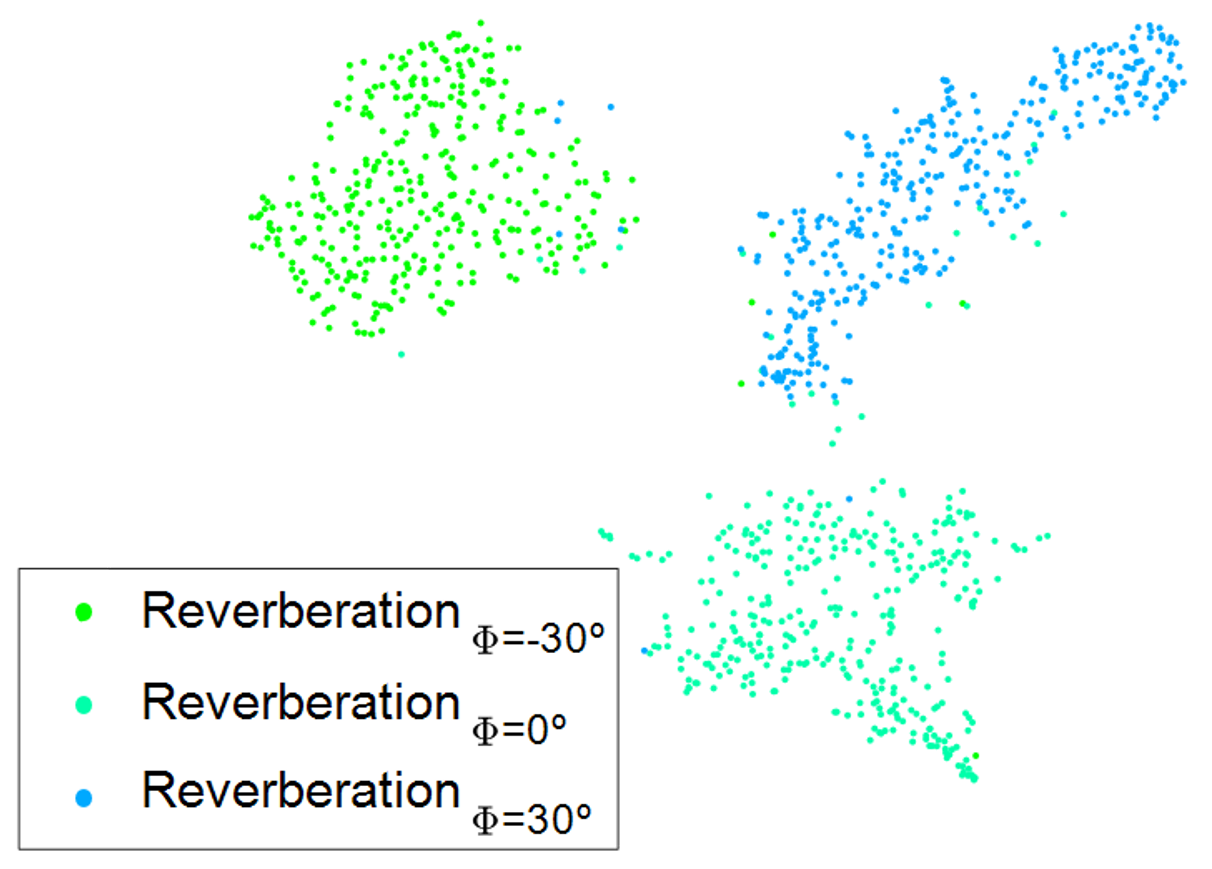}
	\caption{Clustering of the reverberation scenario for three azimuth angles ($-30\degree$, $0\degree$ and $30\degree$). Learning rate and perplexity are fixed to 700 and 90.} 
	\label{cluster_rever}
\end{figure}
%%%%%%%%%%%%%%%%%%%%%%%%%%%%%%%%%%%%%%%%%%%%%%%

%%%%%%%%%%%%%%%%%%%%%%%%%%%%%%%%%
\subsection{\label{sec:Results_B}Deep analysis of Anechoic, Reverberation and Indoor scenarios}
%%%%%%%%%%%%%%%%%%%%%%%%%%%%%%%%%

Since Fig. 5 shows that intra class differences exist, t-SNE is computed for each scenario (1089 datapoints) in order to find intrapopulation separation evidences. The first case is displayed in Fig. 6, where the clustering of the anechoic scenario for three azimuth angles is performed. Two conclusions can be drawn. First, the measurements for $\phi = -30\degree$ and $\phi = 30\degree$ are mixed on the right side of the figure. This fact can be explained from the symmetry of the semi-anechoic chamber in the YZ plane (see Fig. 1). Due to the presence of absorbers, the MPCs are attenuated, and the main MPC (LoS case) reaches \blue{the} RX identically for a positive or negative azimuth angle. Since no information about the angle of arrival (AoA) is included in $\mathbf{x}_i$ [eq. (5)], it is reasonable that both angles are mixed. Secondly, the measurements for $\phi = 0\degree$ tend to be correctly separated from the two previous angles. For this angle, the perfect alignment for the TX-RX antennas causes lower values for the path loss, and therefore, higher spectral efficiency values. These differences between angles explain the large values of standard deviation for the spectral efficiency in Fig. 10(a).

%%%%%%%%%%%%%%%%%%%%%%%%%%%%%%%%%%%%%%%%%%%%%%%
\begin{figure}[t]
	\centering
	\includegraphics[width= 0.9\columnwidth]{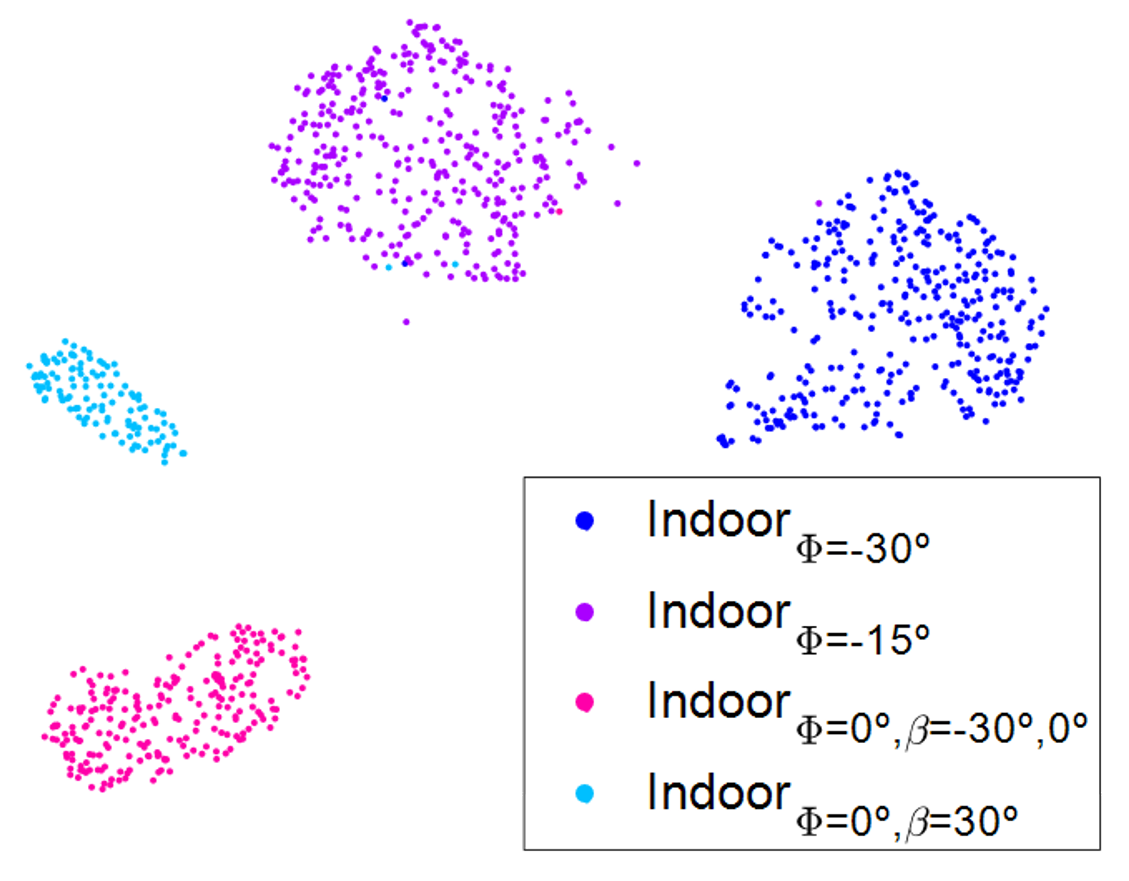}
	\caption{Clustering of the indoor scenario for three azimuth angles ($0\degree$, $-15\degree$ and $-30\degree$) and three roll angles ($-30\degree$, $0\degree$ and $30\degree$) for $\phi=0\degree$. Learning rate and perplexity are fixed to 600 and 140.} 
	\label{cluster_indoor}
\end{figure}
%%%%%%%%%%%%%%%%%%%%%%%%%%%%%%%%%%%%%%%%%%%%%%%

The clustering for the reverberation scenario with three $\phi$ angles is shown in Fig. 7. Although the cases  $\phi = -30\degree$ and  $\phi = 30\degree$ were mixed in Fig. 5, it can be observed a good separation between both angles in Fig. 7. Considering the symmetry of the semi-reverberation chamber, these results may appear to be incorrect.  However, the presence of the chamber door for $\phi = 30\degree$ has to be taken into account, as well as the measuring table [see Fig. 2(a)]. The door and the measuring table break the symmetry of the semi-reverberation scenario in the YZ plane, altering the MPCs. This leads us to conclude that the global dataset tends to \blue{discover} global differences \blue{among} classes (inter class), giving less importance to intra class differences. The input data must be adjusted to the specific classes when intra class differences are searched, as depicted in Fig. 7.

The third case is shown in Fig. 8, where three azimuth angles are clustered for the indoor environment. As shown by the trend in Fig. 5, the three subscenarios are separable into clusters. It is important to note that these three angles correspond to three completely different pointing angles: $\phi = 30\degree$ (TX points \blue{at} the window of the reverberation chamber), $\phi = -15\degree$ (TX points \blue{at} the chamber door frame), and $\phi = 0\degree$ (LoS through the chamber door).  Another interesting detail is the separation of the angle $\phi = 0\degree$ into two subclusters. It has been found that the upper left cluster corresponds to $\beta = 30\degree$, while the lower left cluster groups the datapoints corresponding to angles $\beta = -30\degree$ and $\beta = 0\degree$. Therefore, it is proved that this clustering technique can also separate measurements as a function of the polarization of the incident wave. Note that the clustering of this roll angle implies that the datapoints have been previously separated according to scenario and azimuth angle. As a result, Fig. 8 is the consequence of a clustering in several layers since the roll angle is three-levels depth (scenario, $\phi$ and $\beta$). This fact proves t-SNE's potential to cluster communication scenarios at several depth levels.

%%%%%%%%%%%%%%%%%%%%%%%%%%%%%%%%%
\subsection{\label{sec:Results_C}Anechoic, Reverberation, Indoor, Rooftop and Auditorium}
%%%%%%%%%%%%%%%%%%%%%%%%%%%%%%%%%

The last example mixes the clustering of previously analyzed environments (anechoic, reverberation and indoor) with real communication scenarios (rooftop and auditorium). The clustering from these five environments is shown in Fig. 9. From an analytical viewpoint, it is observed an excellent separation between the three previous scenarios and new (rooftop and auditorium) scenarios. However, rooftop and auditorium seem to be totally mixed. The frequency, the distance and the antenna positions are identical in both scenarios, being the environment surrounding the radiating elements the only difference. In the rooftop scenario, it is expected to obtain a single peak in the PDP due to the outdoor nature. In the auditorium scenario, we would expect to get the main MPC due to the LoS and several reflections from the walls. However, due to the frequency and the distance to the walls of the antennas, the LoS is dominant over any reflection (that is attenuated) and a single peak is seen on the PDP. Therefore, the auditorium scenario behavior is identical to the rooftop scenario. Actually, Figs. 10(d) and 10(e) show how close the communication parameters are for both scenarios, affirming the previous explanation. Therefore, both scenarios are similar in terms of electromagnetic wave propagation. This fact explains that both scenarios are grouped in the same cluster.

%%%%%%%%%%%%%%%%%%%%%%%%%%%%%%%%%%%%%%%%%%%%%%%
\begin{figure}[t]
	\centering
	\includegraphics[width= 0.9\columnwidth]{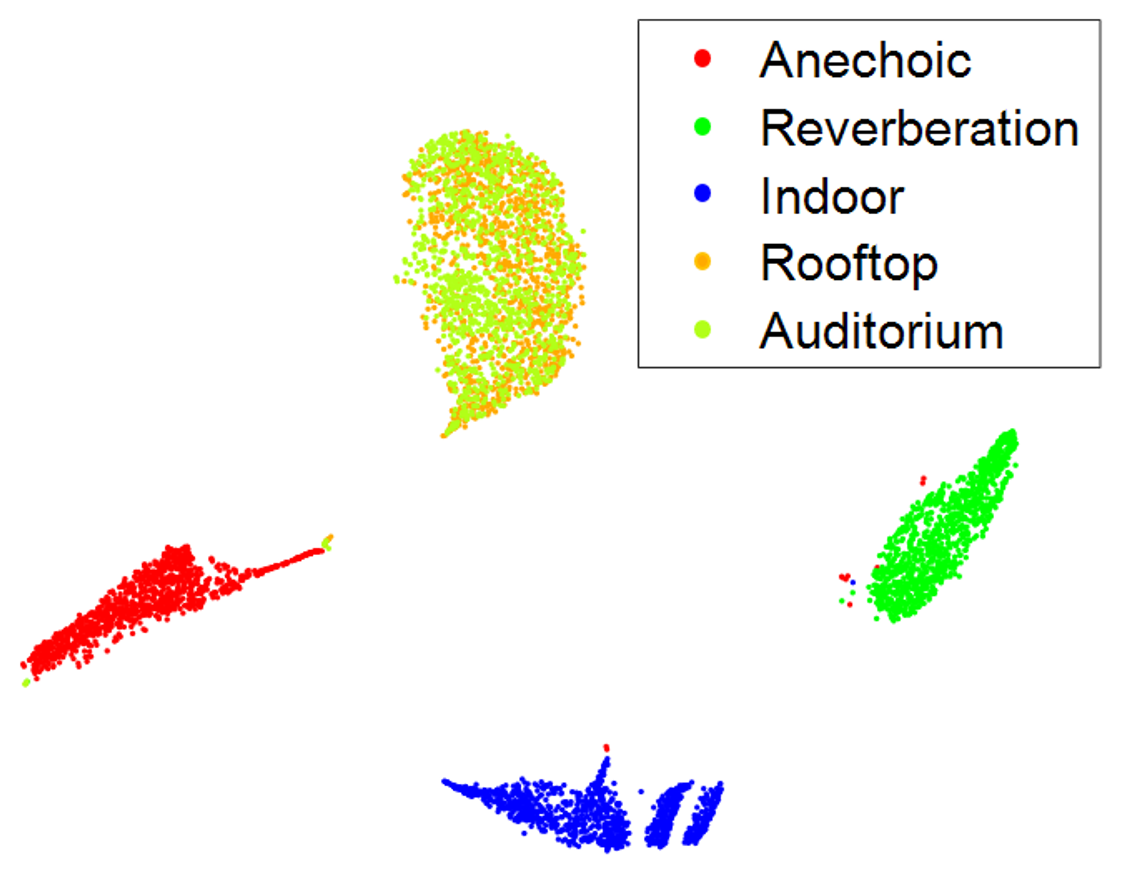}
	\caption{Clustering of the anechoic, reverberation, indoor, rooftop and auditorium scenarios. Learning rate and perplexity are fixed to 650 and 400.} 
	\label{cluster_cinco}
\end{figure}
%%%%%%%%%%%%%%%%%%%%%%%%%%%%%%%%%%%%%%%%%%%%%%%

%%%%%%%%%%%%%%%%%%%%%%%%%%%%%%%%%%%%%%%%%%%%%%%%%%
\begin{figure*}[!t]
    \centering
    \subfigure[]{\includegraphics[width= 0.31\textwidth]{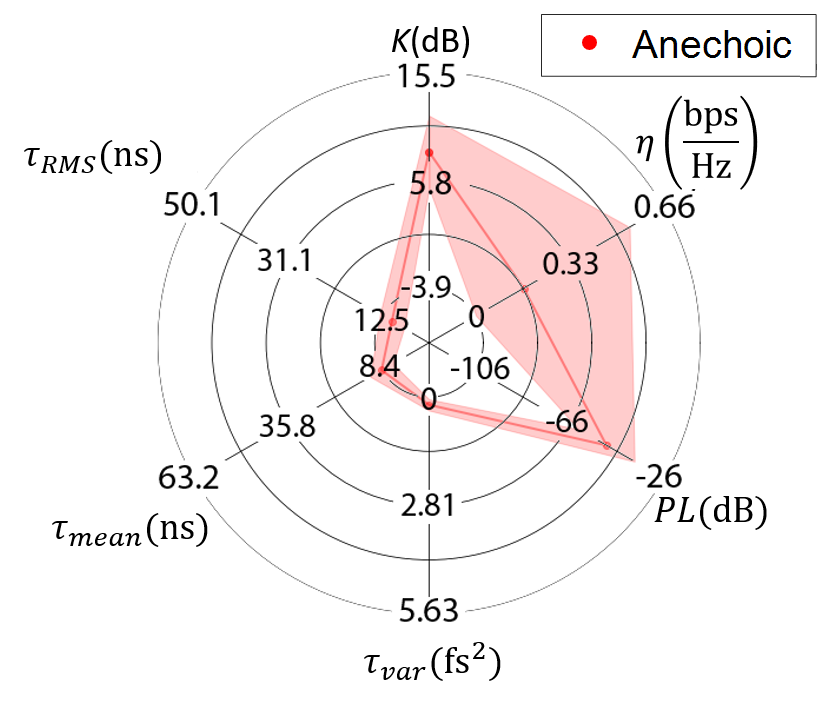}
	}
	\subfigure[]{\includegraphics[width= 0.34\textwidth]{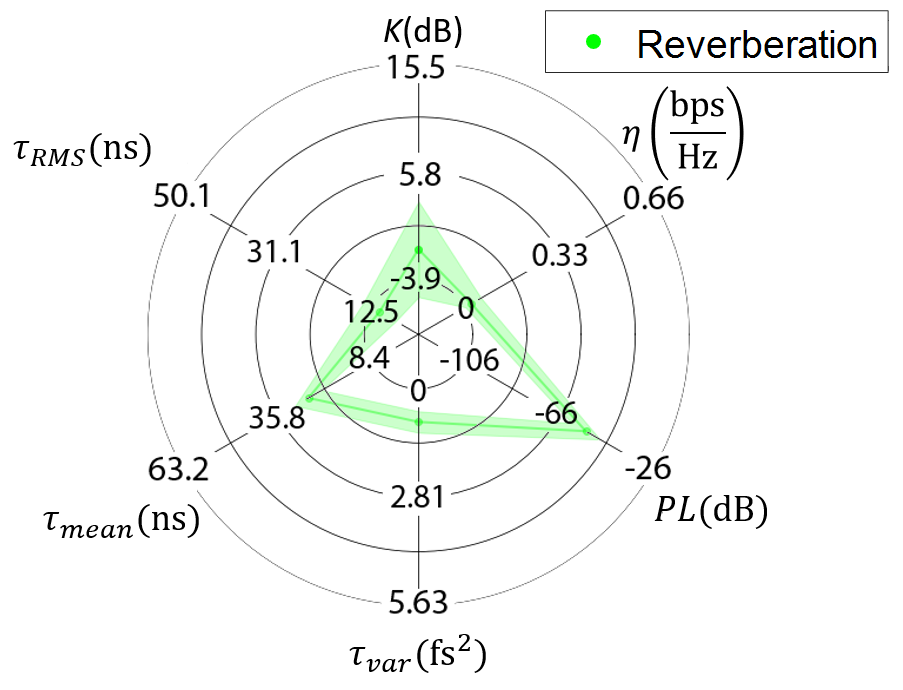}
	}
	\subfigure[]{\includegraphics[width= 0.305\textwidth]{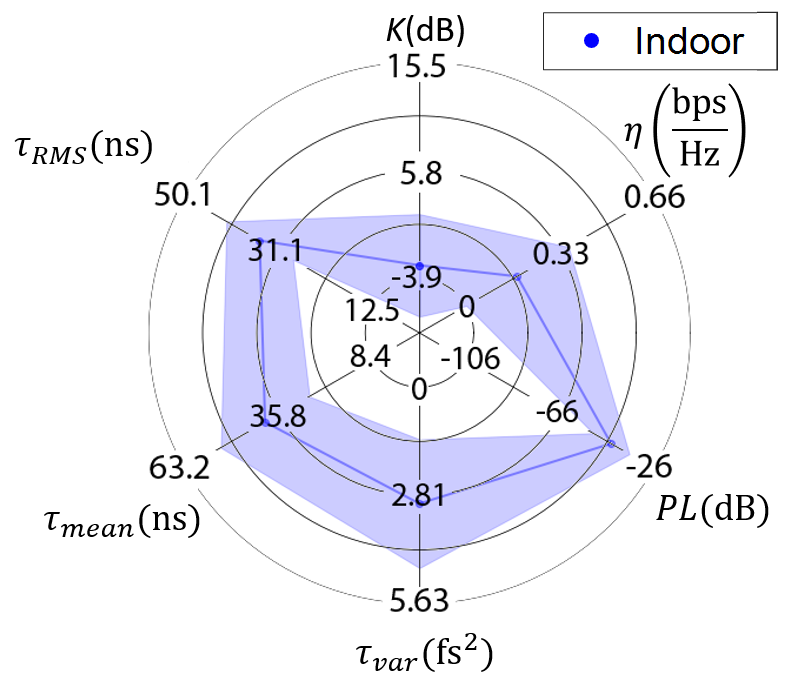}
	}
	\subfigure[]{\includegraphics[width= 0.315\textwidth]{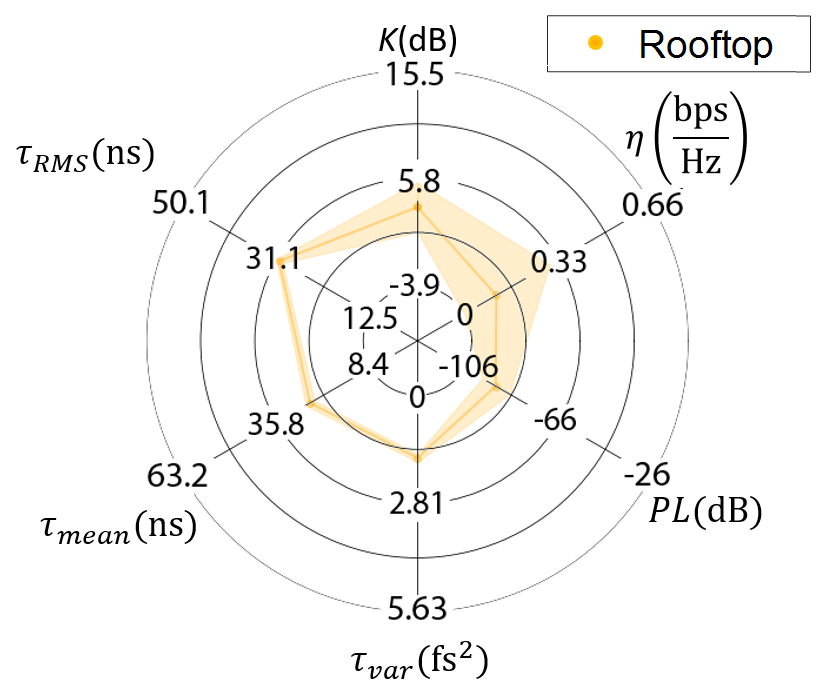}
	}
	\subfigure[]{\includegraphics[width= 0.33\textwidth]{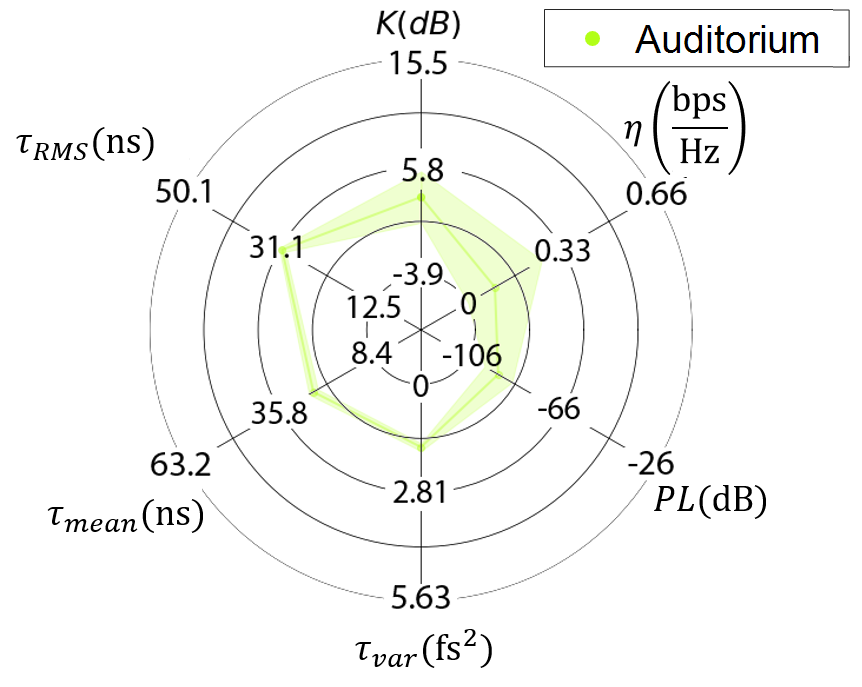}
	}
    \caption{Radar chart of the anechoic (a), reverberation (b), indoor (c), auditorium (d) and rooftop (e) scenarios. The average value for each one of the five datasets is shown in a solid dot. The standard deviation is represented as the width of the lines.} 
	\label{diagrama_arana}
\end{figure*}
%%%%%%%%%%%%%%%%%%%%%%%%%%%%%%%%%%%%%%%%%%%%%%%%%%%

As a last note, it should be remarked how part of the anechoic datapoints tends to approach the rooftop and auditorium data. This data subset represents the orange dots seen earlier in Figs. 5 and 6. These dots, belonging to the case $\phi = 0\degree$, are characterized by a single peak in the time domain corresponding to the LoS path. Since the signal shape in the time domain is similar to the rooftop and auditorium cases, these dots tend to stay close. The difference between them is mainly due to the power amplitude and delay of the signal. This fact opens a way to the rooftop and auditorium recreation from a controlled scenario as the anechoic case. 

%\red{Uno de los revisores comenta: \textit{"A datapoint consists of 6 dimensions as proposed in the paper. Which dimension contributes more to the classification results in your simulation section?} ¿Sabemos qué dimensiones aportan más y menos información para determinar las características de cada escenario? Es interesante comentarlo, ya que futuros trabajos a lo mejor solo necesitan usar 3 o 4 parámetros para llegar a las mismas conclusiones que nosotros, simplificando el estudio. De hecho, esto sería una cosa interesante a destacar en las conclusiones"} 

%%%%%%%%%%%%%%%%%%%%%%%%%%%%%%%%%%%%%%%%%%%%%%%%%%
\begin{figure*}[t]
    \centering
    \includegraphics[width=1\textwidth]{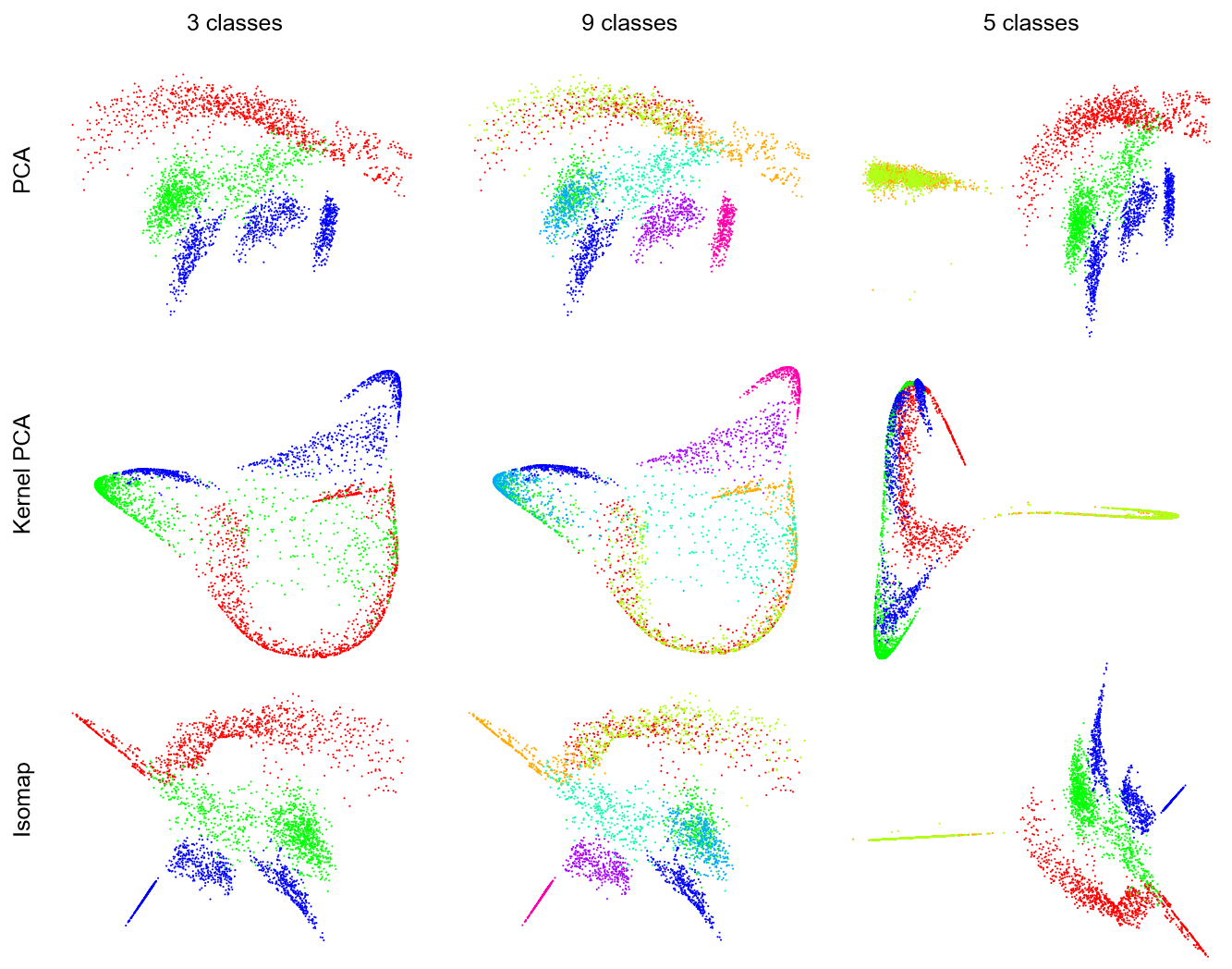}
    \caption{2D dimensionality reduction comparison for multiple techniques and datasets. The rows show the following techniques: PCA, Kernel PCA and Isomap. The columns represent the three datasets used during the work. For comparison with t-SNE, the low-dimensionality spaces in the first, second and third column are equivalent to Figs. 4, 5 and 9, respectively.} 
	\label{comparativa_completa}
\end{figure*}

%%%%%%%%%%%%%%%%%%%%%%%%%%%%%%%%%%%%%%%%%%%%%%%%%%

%%%%%%%%%%%%%%%%%%%%%%%%%%%%%%%%%
\section{\label{sec:Comparacion} Comparing the Performance of Dimensionality Reduction Techniques}
%%%%%%%%%%%%%%%%%%%%%%%%%%%%%%%%%

Throughout the present work, clustering and dimensionality reduction of the communication channels have been performed using the t-SNE technique. Now, the efficiency of t-SNE is compared with other state-of-the-art techniques \blue{in order to demonstrate that this technique shows the best performance of the presented approaches.}

Three datasets are considered in the comparison. The first one (Fig. 4) is composed by the anechoic, reverberation and indoor measurements. The second one (Fig. 5) includes these three scenarios plus the $\phi$ angle differentiation, for a total of 9 classes. The third one (Fig. 9) is formed by the three original scenarios plus the measurements obtained in the FFH Institut, for a total of 5 classes. With respect to the DR techniques, three state-of-the-art techniques are selected to perform this study: PCA \cite{PCA}, Kernel PCA \cite{Kernel_PCA} and Isomap \cite{Isomap}. Principal Component Analysis is one of the oldest techniques used for DR. In summary, this technique aims to reduce the dimensionality of a given dataset by preserving the variance statistical information. For that purpose, this technique calculates linear functions of the variables contained in the original dataset. \blue{These} new variables, known as Principal Components (PCs), are uncorrelated to each other and contain the maximum possible variance from the original dataset. Typically, the PCs are derived by solving a singular value decomposition (see Appendix~A). The analysis of the PCs in our dataset showed that the most influential communication channel parameters for the scenario discrimination are the path loss and the K-factor, followed by the spectral efficiency, and the time parameters ($\tau_{RMS}, \tau_{\text {mean}}, \tau_{\text {var}}$). One step further, Kernel PCA is a PCA extension which allows a nonlinear dimensionality reduction. For that purpose, Kernel PCA applies a nonlinear transformation to the original dataset variables. This is done through the dataset projection with a kernel function, e.g., polynomial, Gaussian, Laplacian \cite{kernel_methods} (see Appendix~A). If the dataset has a structure that can not be separated into a linear subspace, Kernel PCA tends to improve the dimensionality reduction performed by standard PCA. Finally, a technique based on isometric mapping (Isomap) is proposed. The Isomap aim is to preserve the geodesic distance of the dataset in the low-dimensional space. To this end, Isomap generates a neighborhood graph with $K$ neighbors. Therefore, the geodesic distance is estimated as the shorthest paths in the graph where each datapoint is connected to $K$ neighbors. In the end, the dataset is embedded into a low-dimensional space through a eigenvalue decomposition of the matrix formed by geodesic distances (see Appendix~B).

Fig. 11 shows the dimensionality reduction of the datasets previously detailed for PCA, Kernel PCA and Isomap DR techniques. Note that Kernel PCA utilizes a Laplacian kernel with $\gamma = 0.1$ and Isomap uses $K = 15$ neighbors. Visually, the three columns should be compared with the DRs in Figs. 4, 5 and 9, respectively. It can be seen that none of these new proposed techniques improves the performance of t-SNE in terms of visualization. There are several areas in the low-dimensional embedding space where multiple classes are mixed together. Although there is a tendency of separating classes, it is not as evident as we saw in t-SNE.

In order to quantify the quality of these embedding spaces, the fitness function proposed in Section IV [eqs. (20) and (21)] is applied to each low-dimensional space illustrated in Fig. 11. Table I presents the fitness function for several datasets and DR techniques. Note that a higher value of the fitness function is directly related to a better clustering visualization of the considered communication scenarios. Concerning the first and second dataset, t-SNE fitness function outperforms the results obtained by PCA, Kernel PCA and Isomap. For the third dataset, t-SNE also obtains a slightly better value. The results provided in Table I support the visual comparison made in Fig. 11. Both studies show that the performance of t-SNE for communication channel embedding is superior compared to other techniques.

%%%%%%%%%%%%%%%%%%%%%%%%%%%%%%%%%%%%%%%%%%%%%%%%%%%

\renewcommand{\arraystretch}{1.5}
\setlength{\tabcolsep}{10pt}

\begin{table}
    \centering

    \caption {Fitness function for several datasets and dimensionality reduction techniques} \label{tabla_comparativa_fitness} 
    
    \begin{tabular}{c|ccc}
    \hline \hline & 3 classes & 9 classes & 5 classes \\
    \hline PCA & $1.476$ & $2.653$ & $4.184$ \\
    Kernel PCA & $1.620$ & $3.124$ & $2.778$ \\
    %Isomap & $1.508$ & $2.810$ & $4.300$ \\
    t-SNE & $5.234$ & $6.727$ & $4.862$ \\
    \hline \hline
    \end{tabular}

\end{table}

%%%%%%%%%%%%%%%%%%%%%%%%%%%%%%%%%%%%%%%%%%%%%%%%%%%%%%%%%%%

As a final study, a classification of the classes in the low-dimensional space is carried out. If a trained classifier is able to correctly classify the observations in the low-dimensional space, it means that the DR technique properly infers and separates the discriminative aspects of each scenario. Therefore, those techniques where the classifier provides the highest accuracy are the ones whose low-dimensional space includes the most easily discriminable classes. \blue{Five} well-established classifier families are considered: k-Nearest Neighbors (k-NN) \cite{kNN}, Support Vector Machine (SVM) \cite{SVM}, Naive Bayes \cite{NaiveBayes}, Bagging \cite{bagging} \blue{and Linear Discriminant Analysis (LDA) \cite{LDA}}. k-NN classifies a given observation based on the class of its closest neighbors, where $k$ is the number of considered neighbors. SVM finds a mapping such that the different classes of the dataset are divided into subspaces separable by a hyperplane with the largest possible margin. The support vectors (SVs) are those observations that bound the hyperplane (see Appendix~C). Naive Bayes is a classifier based on Bayesian statistics where the observation features are considered independent. Under this assumption, Naive Bayes estimates that an observation belongs to a given class with the maximum \textit{a posteriori} decision rule (see Appendix~D). Bagging, also known as bootstrap aggregation, replicates the original dataset into several new learning datasets. These new datasets are trained independently by using independent classifiers (e.g., k-NN or decision trees). Finally, the classification of each independent classifier is aggregated into a single estimation by the majority of the classifier predictions. \blue{Linear Discriminant Analysis assumes that each scenario is modeled by a multivariate Gaussian distribution (see Appendix E). This model leads to a subspace formed by linear decision boundaries. Therefore, each region includes the observations for a given class.}

%%%%%%%%%%%%%%%%%%%%%%%%%%%%%%%%%%%%%%%%%%%%%%%%%%%%%%%%%%%%
%\renewcommand{\arraystretch}{1.5}
%\setlength{\tabcolsep}{4pt}

%\begin{table*}
%    \centering
%
%    \caption {Classification accuracy by using several classifiers for several datasets and dimensionality reduction techniques} \label{tabla_comparativa} 

%    \begin{tabular}{c|ccc|ccc|ccc|ccc}
%    \hline \hline
%    \multicolumn{1}{c|}{} & \multicolumn{3}{c|}{k-NN} & \multicolumn{3}{c|}{SVM} & \multicolumn{3}{c|}{Naive Bayes} & \multicolumn{3}{c}{Bagging} \\
%    \hline \hline
%     & 3 classes & 9 classes & 5 classes & 3 classes & 9 classes & 5 classes & 3 classes & 9 classes & 5 classes & 3 classes & 9 classes & 5 classes \\
     
%    \hline PCA & $97.55 \%$ & $76.86 \%$ & $84.11 \%$ & $94.25 \%$ & $75.76 \%$ & $78.48 \%$ & $92.72 \%$ & $74.56 \%$ & $73.11 \%$ & $97.40 \%$ & $75.24 \%$ & $83.27 \%$ \\
%    Kernel PCA & $93.57 \%$ & $71.90 \%$ & $75.08 \%$ & $86.84 \%$ & $66.05 \%$ & $50.45 \%$ & $72.79 \%$ & $64.56 \%$ & $58.84 \%$ & $93.85 \%$ & $70.77 \%$ & $73.92 \%$ \\
%    Isomap & $97.70 \%$ & $76.22 \%$ & $83.38 \%$ & $96.63 \%$ & $75.85 \%$ & $76.71 \%$ & $95.35 \%$ & $74.99 \%$ & $72.25 \%$ & $97.43 \%$ & $74.96 \%$ & $82.74 \%$ \\
%    t-SNE & $99.85 \%$ & $81.24 \%$ & $84.00 \%$ & $99.85 \%$ & $79.13 \%$ & $82.04 \%$ & $99.85 \%$ & $75.18 \%$ & $82.06 \%$ & $99.94 \%$ & $80.07 \%$ & $83.23 \%$ \\
%    \hline \hline
%    \end{tabular}
    
%\end{table*}
%%%%%%%%%%%%%%%%%%%%%%%%%%%%%%%%%%%%%%%%%%%%%%%%%%%%%%%%%%%%

%%%%%%%%%%%%%%%%%%%%%%%%%%%%%%%%%%%%%%%%%%%%%%%%%%%%%%%%%%%%
\renewcommand{\arraystretch}{1.5}
\setlength{\tabcolsep}{4pt}

\begin{table*}
    \centering

    \caption {Classification accuracy (\%) by using several classifiers for several datasets and dimensionality reduction techniques } \label{tabla_comparativa} 

    \begin{tabular}{c|ccc|ccc|ccc|ccc|ccc}
    \hline \hline
    \multicolumn{1}{c|}{} & \multicolumn{3}{c|}{k-NN} & \multicolumn{3}{c|}{SVM} & \multicolumn{3}{c|}{Naive Bayes} & \multicolumn{3}{c|}{Bagging} & \multicolumn{3}{c}{\blue{LDA}} \\
    \hline \hline
    \blue{\diagbox[width=1.75cm,height=0.5cm]{\scriptsize{\!\!\!\!\!\!\!\!\!\!\vspace*{-0.09cm} Technique}}{\scriptsize{\raisebox{0.1cm}{Classes}\!\!\!\!\!\!}}} & \{3\} & \{9\} & \{5\} & \{3\} & \{9\} & \{5\} & \{3\} & \{9\} & \{5\} & \{3\} & \{9\} & \{5\} & \blue{\{3\}} & \blue{\{9\}} & \blue{\{5\}} \\
     
    % \backslashbox{\tiny{DR technique}}{\tiny{Classes}}
     
    \hline PCA & $97.55$ & $76.86$ & $84.11$ & $94.25$ & $75.76$ & $78.48$ & $92.72$ & $74.56$ & $73.11$ & $97.40$ & $75.24$ & $83.27$ & \blue{$91.09$} & \blue{$71.22$} & \blue{$73.37$} \\
    Kernel PCA & $93.57$ & $71.90$ & $75.08$ & $86.84$ & $66.05$ & $50.45$ & $72.79$ & $64.56$ & $58.84$ & $93.85$ & $70.77$ & $73.92$ & \blue{$72.66$} & \blue{$62.11$} & \blue{$53.43$} \\
    Isomap & $97.70$ & $76.22$ & $83.38$ & $96.63$ & $75.85$ & $76.71$ & $95.35$ & $74.99$ & $72.25$ & $97.43$ & $74.96$ & $82.74$ & \blue{$95.50$} & \blue{$73.64$} & \blue{$72.76$} \\
    t-SNE & $99.85$ & $81.24$ & $84.00$ & $99.85$ & $79.13$ & $82.04$ & $99.85$ & $75.18$ & $82.06$ & $99.94$ & $80.07$ & $83.23$ & \blue{$99.85$} & \blue{$75.66$} & \blue{$81.03$} \\
    \hline
    \blue{Original 6D Space} & \blue{$99.94$} & \blue{$88.06$} & \blue{$87.16$} & \blue{$99.88$} & \blue{$86.84$} & \blue{$82.31$} & \blue{$97.36$} & \blue{$83.28$} & \blue{$80.73$} & \blue{$99.94$} & \blue{$90.17$} & \blue{$88.28$} & \blue{$99.88$} & \blue{$84.57$} & \blue{$81.68$} \\
    \hline \hline
    \end{tabular}
    
\end{table*}
%%%%%%%%%%%%%%%%%%%%%%%%%%%%%%%%%%%%%%%%%%%%%%%%%%%%%%%%%%%%

In order to perform the classification with each of the \blue{five} classifiers, a 10-fold cross validation is carried out. Each 10-fold cross validation is iterated 10 times for statistical validity purposes. Therefore, the accuracy is calculated as the average accuracy over the ten iterations of the 10-fold cross validation. This procedure avoids outliers due to a non representative sample choice of the training set in the classifiers. Concerning the configuration parameters for the classifiers: k-NN classifier considers $k = 10$ neighbors. SVM classifier uses $({N_c}^2-N_c)/2$ binary learners, i.e., a one-versus-one strategy. The kernel function is linear, which means that a linear boundary separates the classes. For the Naive Bayes classifier, the probabilities according to Bayes rule are distributed following a Gaussian distribution centered on a feature average value given a class. It uses a one-versus-one strategy, providing the same binary classifier number than in SVM. Finally, the boostrap aggregation classifier is formed by 100 independent decision tree classifiers. Each decision tree includes on average 63\% of the observations, which are randomly chosen from the original dataset.

Table II shows the classification accuracy for these \blue{five} classifiers and three datasets. \blue{The accuracy is defined as the number of predicted scenarios that matches the true class over the total number of predicted scenarios for each class, and the considered accuracy is the average for all iterations.} For a given classification technique and dataset \blue{(columns in Table II)}, t-SNE generally obtains the best accuracy compared with PCA, Kernel PCA and Isomap. This fact implies that t-SNE exploits in a better way the discriminative aspects of each scenario. Although the accuracy for some datasets is almost identical, note that the cluster visualization in the low-dimensional space throughout the work is more straightforward in t-SNE. This fact is supported by the proposed fitness function in eqs. (20) and (21), and Table I. 

\blue{As a final comparison, the classification accuracy of the  original 6D space is shown in Table II. Since no information is lost due to the application of a DR technique, the expected accuracy should be larger than cases where DR is applied. However, if DR works properly over the dataset, the classification accuracy with DR should be close to the classification accuracy without DR. Particularly, t-SNE for 3 classes dataset obtains similar accuracies than 6D space. This implies that t-SNE preserves the information from the 6D space into the two-dimensional space. For 9 and 5 classes datasets, t-SNE obtains a misclassification due to DR of 8.4\% and 2.3\% in average, respectively. In exchange for slightly decreasing classification accuracy, the space complexity decreases from 6D to 2D. An unique case is found for the Naive Bayes classifier, where t-SNE gets better classification accuracies than 6D space for 3 and 5 classes datasets. This is explained by the fact that Naive Bayes assumes independence between the communication channel parameters (see Appendix D). This is not meet in the 6D space and it induces classification errors. However, the large separation achieved by t-SNE in 2D (see Figs. 4 and 9) increases the classification accuracy. Finally, it should be remarked that the misclassification between the 6D space and PCA, Kernel PCA and Isomap is larger than in t-SNE, which implies increased information losses in the DR.}

%Aquí comentaría un poco más la importancia de la función de visualización. Es decir, los clasificadores obtienen unos resultados númericos muy parecidos (tabla 2), es decir, clasifican prácticamente igual con PCA, Kernel PCA, pero sin embargo, la visualización es mucho más clara con t-SNE, mirese la visualización de 3 clases (1 y pico) de la figura 13, frente a la figura  4 con un valor de 5.2  

This section has demonstrated by three different experiments (visualization in Fig. 11, fitness function in Table I and accuracy in Table II) that t-SNE provides the \blue{best performance compared to} other state-of-the-art dimensionality reduction techniques.

%%%%%%%%%%%%%%%%%%%%%%%%%%%%%%%%%
\section{\label{sec:Modification} Generation of Scenarios}
%%%%%%%%%%%%%%%%%%%%%%%%%%%%%%%%%

In our previous works \cite{time_gating},\cite{time_gating_eucap}, \cite{juanfra_1}-\hspace{1sp}\cite{juanfra_3}, we have applied post-processing techniques in order to modify several signals in anechoic and reverberation chambers. For example, a time-domain signal acquired in the semi-anechoic and semi-reverberation chamber can be modified to emulate several environments \cite{time_gating},\cite{time_gating_eucap}. These works illustrate that it is possible to recreate different types of communication scenarios from measurements acquired in the chambers. Now, the combination of post-processing techniques and clustering with DR techniques allows the generation and comparison of new communication environments.

In Section IV, it was explained that the anechoic scenario behaves similarly as the rooftop and auditorium scenarios due to similar PDP shapes. The main differences between both and what keeps them apart, are the distances between TX-RX antennas and the acquisition frequency. On the one hand, the time of arrival is affected by the TX-RX distance, which is 160 cm for the anechoic case and 800 cm the distance for the rooftop and auditorium. Therefore, for a propagation speed $c$, the time of arrival is 5.33 ns and 26.66 ns respectively (see Fig. 10). On the other hand, the attenuation is affected, again, by the TX-RX distance and the acquisition frequency (25.875 GHz in the anechoic dataset and 60.48 GHz in the rooftop and auditorium). Since the distance and the frequency are higher in real scenarios, it is logical to find higher path losses in the rooftop and auditorium in Fig. 10.

%%%%%%%%%%%%%%%%%%%%%%%%%%%%%%%%%%%%%%%%%%%%%%%
\begin{figure}[t]
	\centering
	\includegraphics[width= 0.85\columnwidth]{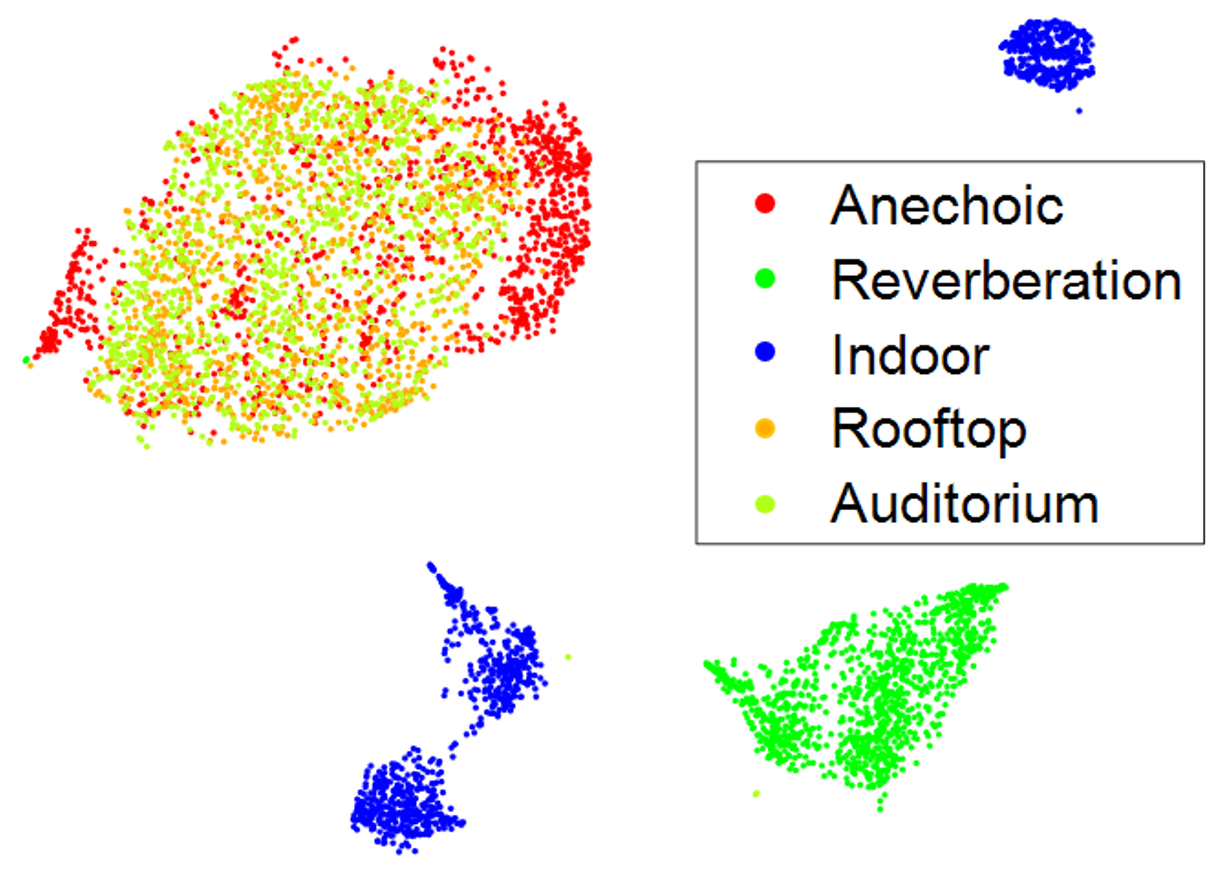}
	\caption{Clustering of the modified anechoic, reverberation, indoor, rooftop and auditorium scenarios. Learning rate and perplexity are fixed to 650 and 400.} 
	\label{cluster_cinco_mod}
\end{figure}
%%%%%%%%%%%%%%%%%%%%%%%%%%%%%%%%%%%%%%%%%%%%%%%

Once the theoretical basis is set, we are able to apply a certain delay and attenuation to the anechoic dataset to obtain a modified version as close as possible to the rooftop and auditorium scenarios. In order to do this, we apply in the frequency domain an attenuation factor to the anechoic dataset such that the average path loss is equal to the average path loss in the rooftop and auditorium, i.e., -93 dB. This attenuation factor is directly applied to the scattering parameters acquired in the CFR. In the time domain, 21.33 ns delay is applied in the anechoic dataset, corresponding to the time difference between the time of arrival of both datasets. This value is also the time that a signal needs to travel the distance difference between scenarios, i.e., 640 cm. Once these changes have been implemented, t-SNE is executed with the original datasets together with the anechoic modified version. The results, presented in Fig. 12, show the formation of a large cluster in the upper left corner. This cluster is formed by the anechoic, rooftop and auditorium datapoints, confirming that the modified anechoic datapoints have joined the real scenarios. The fact that t-SNE cannot separate those three scenarios proves that measurements taken in controlled scenarios can recreate measurements from real scenarios when post-processing techniques are applied.

%%%%%%%%%%%%%%%%%%%%%%%%%%%%%%%%%%%%%%%%%%%%%%%
\begin{figure}[t]
	\centering
	\includegraphics[width= 1\columnwidth]{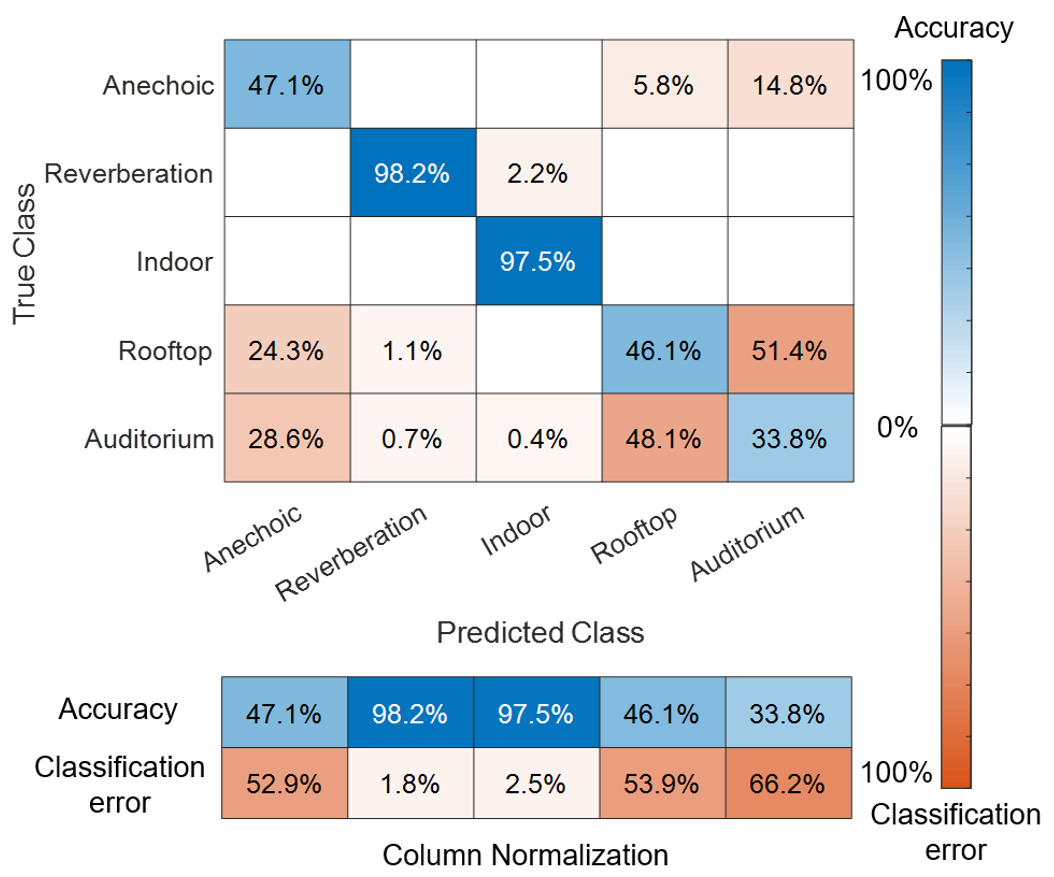}
	\caption{Classification of communication scenarios by using a Variational AutoEncoder.} 
	\label{VAEs}
\end{figure}
%%%%%%%%%%%%%%%%%%%%%%%%%%%%%%%%%%%%%%%%%%%%%%%

To confirm this statement from another perspective, a Variational AutoEncoder (VAE) \cite{VAEs} is applied as classifier to the parameters that constitute the communication channel. VAEs are composed by two neural networks, encoder and decoder, respectively. On the one hand, the encoder decreases the dimensionality of the inputs into a latent space. On the other hand, the decoder learns how to reconstruct the inputs from the latent space. \blue{Therefore, VAEs can be proposed for two different purposes: classification \cite{VAE_clas_1}-\hspace{1sp}\cite{VAE_clas_3} and generation \cite{VAE_gen_1}-\hspace{1sp}\cite{VAE_gen_3}. For classification purposes, the Variational AutoEncoder minimizes the reconstruction error for each class given the latent space. For generation purposes, the encoder provides a set of latent space parameters that characterizes the scenario, and the decoder generates an estimation  $\hat{\mathbf{x}}_{i}$, which should be similar to the original input $\mathbf{x}_{i}$. In this section, the VAE is set as a classifier that assigns one of the five possible classes to each vector $\mathbf{x}_{i}$}. Half of the data from each dataset is chosen to train the VAE, 25\% for validation and 25\% for test. Therefore, 545 datapoints from each class are the input for the training process and 272 datapoints are for validation and test purposes. Fig. 13 shows the classification accuracy of the communication scenarios for the test data once the VAE is trained. As discussed above, indoor and reverberation scenarios are well predicted with accuracy values above 97.5\%. However, the classification of anechoic, rooftop and auditorium scenarios shows poor results with accuracies below 47.1\%.  The average combined accuracy of these three scenarios is 44.24\%. Considering that the classification of the reverberation and indoor cases is good, a random classification between these three scenarios would imply a 33\% accuracy, not far from the accuracy obtained with the VAE. Results in Fig. 13 also confirm that the VAE is not able to optimally separate the modified anechoic environment from the rooftop and auditorium environments. Therefore, the use of post-processing techniques to recreate real scenarios from anechoic measurements is satisfactory. \blue{As future research, we intend to emulate communication scenarios by using the VAEs generator function. Together with the time-gating technique, we expect it to be a powerful generation tool.} This fact, together with the use of generative models and deep reinforcement learning, opens up new possibilities for the generation of future mobile communication scenarios as Vehicle-to-Vehicle, UAV-to-UAV, Ship-to-Ship or High Speed Train-to-High Speed Train.

%%%%%%%%%%%%%%%%%%%%%%%%%%%%%%%%%
\section{\label{sec:Conclusions} Conclusions}
%%%%%%%%%%%%%%%%%%%%%%%%%%%%%%%%%

This work presents a deep analysis of the t-SNE technique to cluster several communication channel scenarios. t-SNE is a well-known technique in the AI and machine learning fields. However, this technique has not been widely exploited in the telecommunication field.  

Six channel features have been extracted from 5089 measurements of five types of communication scenarios: (i) anechoic, (ii) reverberation, (iii) indoor, (iv) rooftop and (v) auditorium. A deep study on the configuration parameters (learning rate and perplexity) has been performed for the six-dimensional space formed by the communication channel features. For fitting configuration values, t-SNE exhibits outstanding performance in the ability to separate communication channel scenarios. Moreover, it has been seen how t-SNE is able to cluster at multiple sublevels, performing a deep clustering. The case shown in Section IV separates the scenario itself and the azimuth and roll angles into subclasses, reaching three-levels depth. In fact, Section V has proved the best performance of t-SNE as DR technique compared with other techniques of the state-of-the-art. \blue{Future research work includes the comparison of t-SNE performance with others non-linear techniques which preserve the data local structure, e.g., Uniform Manifold Approximation and Projection (UMAP).}

Finally, we have modified, by applying suitable post-processing techniques, one of the controlled scenarios in order to recreate another of the real scenarios. The modification of the delay and attenuation of a dataset formed by an anechoic scenario has led to a scenario with similar propagation conditions. The application of the t-SNE technique as a clustering algorithm and a VAE as a classifier has proved that the modified version of the anechoic scenario is confused with the real scenarios. This fact demonstrates that the modification of communication scenarios with post-processing techniques is feasible to recreate new communication environments.

As stated in Section V, although VAEs are used for classification in this work, they have a much bigger potential as generative models once the probability distribution of known scenarios is well \blue{modeled}. Therefore, future work will include the generation of new scenarios based on the knowledge acquired by the VAEs through the clustering and classification analysis. The generation of new communication scenarios seems fundamental from a technology perspective due to the exponential growth of the users and devices in the world. New mobile communication paradigms need a deep understanding of the propagation environments where the communications are held. The clustering technique for communication scenarios and the recreation of scenarios shown throughout this work arise as powerful tools to simplify the \blue{understanding of a telecommunication deployment in terms of the channels through which communications can take place} for future mobile communications. \blue{These tools are able to provide a simple visualization of channel similarities which at first sight may appear to be different and thus increase the knowledge of the communication channels in a telecommunication deployment.}

\begin{appendix}

\section*{A. Principal Component Analysis}

%\red{Algunos vectores y matrices ya están definidos en el texto, no tienes por qué volver a definirlos. Simplemente recuerda a la gente que $\mathbf{X}$ es la matriz de alta dimension y un par de historias más.}

Let $\mathbf{X} \in \mathbb{R}^{N \times F}$ be the matrix that represents the high-dimensional space dataset\blue{, where $\mathbf{X}$ is centralized as $\mathbf{X} \rightarrow \mathbf{X} - \mathbf{\bar{X}}$}. The single value decomposition (SVD) of $\mathbf{X}$ is:

\begin{equation}
    \mathbf{X = P \Delta Q^{\mathit{T}}}
\end{equation}

\noindent where $\mathbf{P} \in \mathbb{R}^{N \times L}$ contains the eigenvectors of the matrix  $\mathbf{X}\mathbf{X^{\mathit{T}}}$ and $\mathbf{Q} \in \mathbb{R}^{F \times L}$ contains the eigenvectors of $\mathbf{X^{\mathsf{T}}}\mathbf{X}$. $\mathbf{\Delta}^2 = \mathbf{\Lambda}$, where $\mathbf{\Lambda}$ is the diagonal matrix that includes the eigenvalues of the matrix $\mathbf{X}\mathbf{X^{\mathit{T}}}$ (and $\mathbf{X^{\mathit{T}}}\mathbf{X}$). Note that $L$ is the rank of $\mathbf{X}$.
We can define $\mathbf{F} \in \mathbb{R}^{N \times L}$ as:

\begin{equation}
    \mathbf{F = P \Delta}
\end{equation}
\noindent where the rows of  $\mathbf{F}$ include the principal components into the low-dimensional space, also called projections. By combining eqs. (22) and (23), it can be seen that

\begin{equation}
    \mathbf{F = XQ}
\end{equation}
\noindent Therefore, $\mathbf{Q}$ can be denoted as the projection matrix, since it provides the linear combination needed to obtain the low-dimensional space. %\red{Note that all the previous steps are usually preceded by the data centralization of the $\mathbf{X}$ matrix ($\mathbf{X} \rightarrow \mathbf{X} - \mathbf{\bar{X}}$).}

For the Kernel PCA technique, we have used a Laplacian transformation as kernel function:

\begin{equation}
\mathbf{K}\left(\mathbf{x}_{i}, \mathbf{x}_{j}\right)=\exp \left(- \gamma \|\mathbf{x}_{i}-\mathbf{x}_{j}\right\|)
\end{equation}
\noindent $\mathbf{K} \in \mathbb{R}^{N \times N}$ is the kernel matrix and $\mathbf{x}_{i}$, $\mathbf{x}_{j} \in \mathbb{R}^{1 \times F}$ are the rows $i$ and $j$ of $\mathbf{X}$. By using this kernel matrix $\mathbf{K}$ together with the kernel trick \cite{Kernel_PCA}\blue{, it} is possible to obtain the low-dimensional space of KPCA without an explicit mapping. This problem could by solved by applying the Laplacian transformation $ \mathbf{X} \rightarrow \Phi\left(\mathbf{X}\right)$, followed by standard PCA [eqs. (22), (23) and (24)]. However, this is not usually performed for computational reasons.

\section*{B. Isomap}

Let $\mathbf{X} \in \mathbb{R}^{N \times F}$ be the high-dimensional space of our dataset and $\mathbf{x}_{i}$, $\mathbf{x}_{j} \in \mathbb{R}^{1 \times F}$ be the rows of $\mathbf{X}$. In the first step, through k-NN, a graph $G$ is created by connecting the node $i$ with the node $j$ if it is one of its closest $K$ neighbors. In the second step, the Dijkstra's algorithm is applied to $G$ in order to find the shortest path between observations $i$ and $j$. These distances are saved in the matrix $\mathbf{D} \in \mathbb{R}^{N \times N}$. In the last step, the low-dimensional space is created by applying multidimensional scaling (MDS) \cite{MDS} to the graph distances~$\mathbf{D}$.

\section*{C. Support Vector Machine}

SVM binary classifier is a supervised learning algorithm which looks for a hyperplane that optimally separates a dataset into two classes. The hyperplane is also intended to maximize the margin between both classes. Two types of classes can take place: separable classes, where the margin between the hyperplanes is not crossed by any observations, and inseparable clases, where the hyperplane can not separate both classes and a penalty is applied for the observations which cross the boundary. Under the inseparable classes assumption, the hyperplane function is given by:

\begin{equation}
f(\mathbf{x}_i)=\mathbf{x}_i\boldsymbol\beta+b=0
\end{equation}

\noindent where $\boldsymbol\beta \in \mathbb{R}^{F \times 1}$ and $b \in \mathbb{R}^{1}$ are the coefficients that define a hyperplane orthogonal vector and a bias term, respectively.

The optimal hyperplane is found by minimizing:

\begin{subequations}
    \begin{align}
      \min \left(\frac{1}{2}  \|\boldsymbol\beta\|^2 + \sum_{i} \kappa_{i}\right) \\
      \text{subject to: } y_{i} f\left(\mathbf{x}_{i}\right) \geq 1-\kappa_{i} \\
      \kappa_{i} \geq 0
    \end{align}
  \end{subequations}

\noindent where $\kappa_{i}$ is a penalty score applied to the $i$-th observation if it crosses the boundary decision defined by the hyperplane. $y_{i} = \{1, -1\}$ stands for the class of the observation.

This optimization problem is typically solved with the method of Lagrange multipliers. The solution defines the coefficients of $\boldsymbol\beta$ and the bias term $b$, which form the boundary decision.

\section*{D. Naive Bayes}

Naive Bayes is a supervised learning algorithm that learns the data distribution through a training set. This algorithm takes advantage of the Bayes rule and uses density estimation in a certain test set. The key point for this classifier is to assume the features (communication channel parameters) to be independent given a class. Therefore, in a binary learner classifier, $2F$ independent predictors are created during the training process. Let $Z{j|k} \sim \mathcal{N}(\mu_{j|k},\,{\sigma_{j|k}}^{2})$ for $j = 1,...,F$ and $k = 1,-1$ be a normal distribution with $\mu_{j|k}$ mean and  $\sigma_{j|k}$ standard deviation for feature $j$ and class $k$. Mean and standard deviation are calculated as:

\begin{equation}
\mu_{j \mid k}=\frac{1}{N_k}\sum_{\left\{i: y_{i}=k\right\}} x_{i j}
\end{equation}

\begin{equation}
\sigma_{j \mid k}=\sqrt{\frac{1}{N_k-1}\sum_{\left\{i: y_{i}=k\right\}} \left(x_{i j}-\mu_{j \mid k}\right)^{2}}
\end{equation}

\noindent where $N_k$ indicates the number of observations of the class $k$ in the training set, and $\sum_{\left\{i: y_{i}=k\right\}}$ is the summation only including the observations that belong to the class $k$.

Once the predictors are trained, the probability that an observation $i$ belongs to the class $k$ given $F$ features is calculated as:

\begin{multline}
\widehat{P}\left(Y_i=k \mid Z_{1},..., Z_{F}\right)= \\ \frac{P_{prior}(Y_i=k) \prod_{j=1}^{F} P\left(Z_{j} \mid Y_i=k\right)}{\sum_{k=1} P_{prior}(Y_i=k) \prod_{j=1}^{F} P\left(Z_{j} \mid Y_i=k\right)}
\end{multline}

\noindent $P_{prior}(Y_i=k)$ is the prior probability that an observation~$i$ belongs to the class $k$, calculated as the density of each class in the training set. Note that $\prod_{j=1}^{F} P\left(Z_{j} \mid Y_i=k\right)$ can be performed due to the feature independence assumption. 

Finally, the $i$-th observation is assigned to the class that generates the maximum \textit{a posteriori} probability.

\section*{\blue{E. Linear Discriminant Analysis}}

\blue{Discriminant Analysis is a supervised learning algorithm that divides a feature space into regions. LDA implies that these regions are linearly separated, where each region includes an observation set which belongs to a certain class $k$. Mathematically, the observation $\mathbf{x}_{i}$ is assigned to the class $k$ according to the classifier $y_{i}$, where $\widehat{y}_{i}$ estimation is calculated as:}
\begin{equation}
\blue{\widehat{y}_{i}=\underset{y_{i}=1, \ldots, K}{\arg \min } \sum_{k=1, k \neq y_{i}}^{K} \widehat{P}(k \mid \mathbf{x}_{i})}
\end{equation}

\noindent\blue{$\widehat{P}(k \mid \mathbf{x}_{i})$ is the estimated probability of belonging to class $k$, given the observation $\mathbf{x}_{i}$. It is calculated as:}

\begin{equation}
\blue{\widehat{P}(k \mid \mathbf{x}_{i})= \frac{P(k)}{P(\mathbf{x}_{i})} \, P(\mathbf{x}_{i} \mid k ) }
\end{equation}

\noindent\blue{where $P(\mathbf{x}_{i} \mid k )$ is a $F-$dimensional multivariate normal density function for the observation $\mathbf{x}_{i}$ given the class $k$:}

\begin{equation}
\blue{P(\mathbf{x}_{i} \mid k ) = 
\frac{1}{\sqrt{(2 \pi)^{F}\left|\xi\right|}} e^{ \left(-\frac{1}{2}\left(\mathbf{x}_{i}-\mathbf{\mu}_{k}\right) \xi^{-1}\left(\mathbf{x}_{i}-\mathbf{\mu}_{k}\right)^{\mathit{T}}\right)}}
\end{equation}

\noindent\blue{$\xi\in \mathbb{R}^{F \times F}$ is the covariance matrix of $\mathbf{X}$ and $\mathbf{\mu}_{k} \in \mathbb{R}^{1 \times F}$ is the mean for each feature given the class $k$. Both are computed through the training dataset. Finally, $P(k)$ is the class distribution on the training set and $P(\mathbf{x}_{i})$ is a normalization factor calculated as $\sum_{k} P(\mathbf{x}_{i} \mid k) \cdot P(k)$.}

\blue{The $i$-th observation is assigned to the class $k$ that minimizes the expected cost in eq. (31).}

%\begin{equation}
%\widehat{P}(k \mid x)=\frac{P(x \mid k) P(k)}{P(x)}
%\end{equation}

\end{appendix}

%%%%%%%%%%%%%%%%%%%%%%%%%%%%%%%
\section*{Acknowledgment}
%\textcolor{red}{Incluir Agradecimientos}

The authors would like to thank the Fraunhofer-Heinrich-Hertz-Institut for acquiring and sharing the data associated to the rooftop and auditorium communication scenarios, the NextG Channel Model Alliance for creating a space to share public databases of propagation measurements, José Francisco Cortés-Gómez for the graphical support, Carmelo García-García for his help in the measurements acquisition, and Sohrab Vafa, Pablo Padilla and Francisco Luna-Valero for their valuable comments.

\ifCLASSOPTIONcaptionsoff
  \newpage
\fi

\end{document}